\pdfoutput=1
\documentclass[12pt,a4paper]{article}
\usepackage{ifthen} % for conditional statements
\usepackage{xcolor}
\newboolean{pdflatex}
\setboolean{pdflatex}{true} % False for eps figures 

\newboolean{articletitles}
\setboolean{articletitles}{true} % False removes titles in references

\newboolean{uprightparticles}
\setboolean{uprightparticles}{false} %True for upright particle symbols

\newboolean{inbibliography}
\setboolean{inbibliography}{false} %True once you enter the bibliography

\usepackage{microtype}
\usepackage{lineno}  % for line numbering during review
\usepackage{xspace} % To avoid problems with missing or double spaces after
\usepackage{caption} %these three command get the figure and table captions automatically small

\usepackage{multirow}
\usepackage{verbatim}
\usepackage{graphicx}  % to include figures (can also use other packages)
\usepackage{color}
\usepackage{colortbl}

\usepackage{amsmath} % Adds a large collection of math symbols
\usepackage{amssymb}
\usepackage{amsfonts}
\usepackage{upgreek} % Adds in support for greek letters in roman typeset

\newcommand*\patchAmsMathEnvironmentForLineno[1]{%
\expandafter\let\csname old#1\expandafter\endcsname\csname #1\endcsname
\expandafter\let\csname oldend#1\expandafter\endcsname\csname
end#1\endcsname
 \renewenvironment{#1}%
   {\linenomath\csname old#1\endcsname}%
   {\csname oldend#1\endcsname\endlinenomath}%
}
\newcommand*\patchBothAmsMathEnvironmentsForLineno[1]{%
  \patchAmsMathEnvironmentForLineno{#1}%
  \patchAmsMathEnvironmentForLineno{#1*}%
}
\AtBeginDocument{%
\patchBothAmsMathEnvironmentsForLineno{equation}%
\patchBothAmsMathEnvironmentsForLineno{align}%
\patchBothAmsMathEnvironmentsForLineno{flalign}%
\patchBothAmsMathEnvironmentsForLineno{alignat}%
\patchBothAmsMathEnvironmentsForLineno{gather}%
\patchBothAmsMathEnvironmentsForLineno{multline}%
}

\usepackage{hyperref}    % Hyperlinks in references
\usepackage[all]{hypcap} % Internal hyperlinks to floats.

\def\lhcb {\mbox{LHCb}\xspace}
\def\cdf    {\mbox{CDF}\xspace}
\def\babar  {\mbox{BaBar}\xspace}
\def\belle  {\mbox{Belle}\xspace}
\ifthenelse{\boolean{uprightparticles}}%
{

 \def\Pmu         {\ensuremath{\upmu}\xspace}                 
 \def\Pnu         {\ensuremath{\upnu}\xspace}                 
                  
 \def\Ppi         {\ensuremath{\uppi}\xspace}

 \def\Ppsi        {\ensuremath{\uppsi}\xspace}

 \def\PDelta      {\ensuremath{\Delta}\xspace}                 
 \def\PXi      {\ensuremath{\Xi}\xspace}                 
 \def\PLambda      {\ensuremath{\Lambda}\xspace}                 
 \def\PSigma      {\ensuremath{\Sigma}\xspace}                 
 \def\POmega      {\ensuremath{\Omega}\xspace}                 
 \def\PUpsilon      {\ensuremath{\Upsilon}\xspace}                 
 
 \def\PB      {\ensuremath{\mathrm{B}}\xspace}                 
                  
 \def\PD      {\ensuremath{\mathrm{D}}\xspace}

 \def\PJ      {\ensuremath{\mathrm{J}}\xspace}                 
 \def\PK      {\ensuremath{\mathrm{K}}\xspace}

 \def\Pb      {\ensuremath{\mathrm{b}}\xspace}                 
 \def\Pc      {\ensuremath{\mathrm{c}}\xspace}

 \def\Pi      {\ensuremath{\mathrm{i}}\xspace}

 \def\Pp      {\ensuremath{\mathrm{p}}\xspace}

 \def\Ps      {\ensuremath{\mathrm{s}}\xspace}

}
{

 \def\Pmu         {\ensuremath{\mu}\xspace}                 
 \def\Pnu         {\ensuremath{\nu}\xspace}                 
                  
 \def\Ppi         {\ensuremath{\pi}\xspace}

 \def\Ppsi        {\ensuremath{\psi}\xspace}                 
                  
 \mathchardef\PDelta="7101
 \mathchardef\PXi="7104
 \mathchardef\PLambda="7103
 \mathchardef\PSigma="7106
 \mathchardef\POmega="710A
 \mathchardef\PUpsilon="7107
                  
 \def\PB      {\ensuremath{B}\xspace}                 
                  
 \def\PD      {\ensuremath{D}\xspace}

 \def\PJ      {\ensuremath{J}\xspace}                 
 \def\PK      {\ensuremath{K}\xspace}

 \def\Pb      {\ensuremath{b}\xspace}                 
 \def\Pc      {\ensuremath{c}\xspace}

 \def\Pi      {\ensuremath{i}\xspace}

 \def\Pp      {\ensuremath{p}\xspace}

 \def\Ps      {\ensuremath{s}\xspace}

}
\def\proton      {{\ensuremath{\Pp}}\xspace}
\def\Lz          {{\ensuremath{\PLambda}}\xspace}  
 
\def\mup        {{\ensuremath{\Pmu^+}}\xspace}
\def\neu        {{\ensuremath{\Pnu}}\xspace}

\def\neum       {{\ensuremath{\neu_\mu}}\xspace}

\def\squark    {{\ensuremath{\Ps}}\xspace}
\def\squarkbar {{\ensuremath{\overline \squark}}\xspace}
\def\ssbar     {{\ensuremath{\squark\squarkbar}}\xspace}
\def\cquark    {{\ensuremath{\Pc}}\xspace}
\def\cquarkbar {{\ensuremath{\overline \cquark}}\xspace}
\def\ccbar     {{\ensuremath{\cquark\cquarkbar}}\xspace}
\def\bquark    {{\ensuremath{\Pb}}\xspace}
\def\bquarkbar {{\ensuremath{\overline \bquark}}\xspace}

\def\pion   {{\ensuremath{\Ppi}}\xspace}

\def\pip    {{\ensuremath{\pion^+}}\xspace}
\def\pim    {{\ensuremath{\pion^-}}\xspace}

\def\kaon    {{\ensuremath{\PK}}\xspace}
  \def\Kbar    {{\kern 0.2em\overline{\kern -0.2em \PK}{}}\xspace}

\def\Kp      {{\ensuremath{\kaon^+}}\xspace}
\def\Km      {{\ensuremath{\kaon^-}}\xspace}

\def\Kstarz  {{\ensuremath{\kaon^{*0}}}\xspace}
\def\Kstarzb {{\ensuremath{\Kbar^{*0}}}\xspace}

  \def\Dbar    {{\kern 0.2em\overline{\kern -0.2em \PD}{}}\xspace}
\def\D       {{\ensuremath{\PD}}\xspace}

\def\Dzb     {{\ensuremath{\Dbar^0}}\xspace}

\def\Dstarm  {{\ensuremath{\D^{*-}}}\xspace}

\def\Dsp     {{\ensuremath{\D^+_\squark}}\xspace}
\def\Dsm     {{\ensuremath{\D^-_\squark}}\xspace}

\def\B       {{\ensuremath{\PB}}\xspace}
\def\Bbar    {{\ensuremath{\kern 0.18em\overline{\kern -0.18em \PB}{}}}\xspace}

\def\Bu      {{\ensuremath{\B^+}}\xspace}

\def\Bp      {{\ensuremath{\Bu}}\xspace}

\def\Bd      {{\ensuremath{\B^0}}\xspace}
\def\Bs      {{\ensuremath{\B^0_\squark}}\xspace}
\def\Bsb     {{\ensuremath{\Bbar^0_\squark}}\xspace}

\def\jpsi     {{\ensuremath{{\PJ\mskip -3mu/\mskip -2mu\Ppsi\mskip 2mu}}}\xspace}

  \def\Y#1S{\ensuremath{\PUpsilon{(#1S)}}\xspace}% no space before {...}!

\def\Lb      {{\ensuremath{\Lz^0_\bquark}}\xspace}

\def\CP                {{\ensuremath{C\!P}}\xspace}

\def\T               {{\ensuremath{T}}\xspace}

\newcommand{\dms}{{\ensuremath{\Delta m_{\squark}}}\xspace}

\newcommand{\DGs}{{\ensuremath{\Delta\Gamma_{\squark}}}\xspace}

\newcommand{\Gs}{{\ensuremath{\Gamma_{\squark}}}\xspace}

\newcommand{\decay}[2]{\ensuremath{#1\!\to #2}\xspace}         % {\Pa}{\Pb \Pc}
\def\BsPP      {\decay{\Bs}{\phi\phi}}

\newcommand{\tev}{\ensuremath{\mathrm{\,Te\kern -0.1em V}}\xspace}
\newcommand{\gev}{\ensuremath{\mathrm{\,Ge\kern -0.1em V}}\xspace}
\newcommand{\mev}{\ensuremath{\mathrm{\,Me\kern -0.1em V}}\xspace}
\newcommand{\kev}{\ensuremath{\mathrm{\,ke\kern -0.1em V}}\xspace}
\newcommand{\ev}{\ensuremath{\mathrm{\,e\kern -0.1em V}}\xspace}
\newcommand{\gevc}{\ensuremath{{\mathrm{\,Ge\kern -0.1em V\!/}c}}\xspace}
\newcommand{\mevc}{\ensuremath{{\mathrm{\,Me\kern -0.1em V\!/}c}}\xspace}
\newcommand{\gevcc}{\ensuremath{{\mathrm{\,Ge\kern -0.1em V\!/}c^2}}\xspace}
\newcommand{\gevgevcccc}{\ensuremath{{\mathrm{\,Ge\kern -0.1em V^2\!/}c^4}}\xspace}
\newcommand{\mevcc}{\ensuremath{{\mathrm{\,Me\kern -0.1em V\!/}c^2}}\xspace}

\def\mum  {\ensuremath{{\,\upmu\rm m}}\xspace}

\def\invfb   {\ensuremath{\mbox{\,fb}^{-1}}\xspace}

\def\ps   {\ensuremath{{\rm \,ps}}\xspace}
\def\fs   {\ensuremath{\rm \,fs}\xspace}

\def\invps{\ensuremath{{\rm \,ps^{-1}}}\xspace}

\newcommand{\stat}{\ensuremath{\mathrm{\,(stat)}}\xspace}
\newcommand{\syst}{\ensuremath{\mathrm{\,(syst)}}\xspace}

\newcommand{\chisq}{\ensuremath{\chi^2}\xspace}

\newcommand{\chisqip}{\ensuremath{\chi^2_{\rm IP}}\xspace}

\def\deriv {\ensuremath{\mathrm{d}}}

\def\gsim{{~\raise.15em\hbox{$>$}\kern-.85em
          \lower.35em\hbox{$\sim$}~}\xspace}
\def\lsim{{~\raise.15em\hbox{$<$}\kern-.85em
          \lower.35em\hbox{$\sim$}~}\xspace}

\def\sPlot{{\em sPlot}\xspace}

\def\pt         {\mbox{$p_{\rm T}$}\xspace}

\def\rad{\ensuremath{\rm \,rad}\xspace}

\def\evtgen     {\mbox{\textsc{EvtGen}}\xspace}

\def\gauss      {\mbox{\textsc{Gauss}}\xspace}
\def\geant      {\mbox{\textsc{Geant4}}\xspace}

\def\photos     {\mbox{\textsc{Photos}}\xspace}

\def\pythia     {\mbox{\textsc{Pythia}}\xspace}

\def\tell1  {TELL1\xspace}
\def\ukl1   {UKL1\xspace}

\newcommand{\phisPP}{\ensuremath{\phi_s}\xspace}
\newcommand{\mFk}{\ensuremath{m_\Kp\kern-0.1 \Km\kern-0.1 \Kp\kern-0.1 \Km}\xspace}
\newcolumntype{R}{>{\raggedleft\arraybackslash}p{1.3cm}}
\newcolumntype{S}{>{\raggedleft\arraybackslash}p{1.3cm}}

\usepackage{cite} % Allows for ranges in citations
\usepackage{mciteplus}

\begin{document}

%%%%%%%%%%%%%%%%%%%%%%%%%
%%%%% Title     %%%%%%%%%
%%%%%%%%%%%%%%%%%%%%%%%%%
\renewcommand{\thefootnote}{\fnsymbol{footnote}}
\setcounter{footnote}{1}

\begin{titlepage}
\pagenumbering{roman}

% Header ---------------------------------------------------
\vspace*{-1.5cm}
\centerline{\large EUROPEAN ORGANIZATION FOR NUCLEAR RESEARCH (CERN)}
\vspace*{1.5cm}
\hspace*{-0.5cm}
\begin{tabular*}{\linewidth}{lc@{\extracolsep{\fill}}r}
\ifthenelse{\boolean{pdflatex}}% Logo format choice
{\vspace*{-2.7cm}\mbox{\!\!\!\includegraphics[width=.14\textwidth]{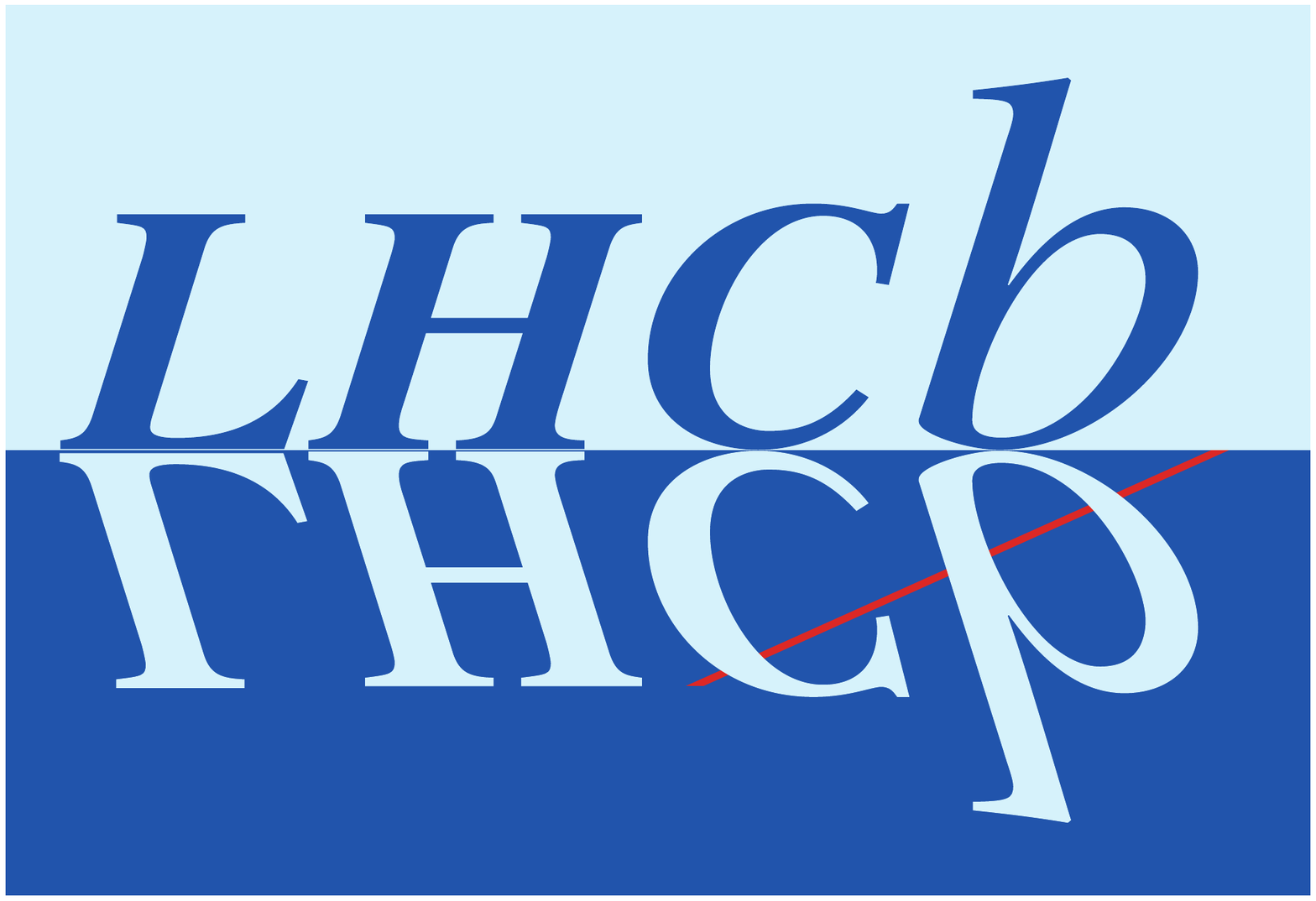}} & &}%
{\vspace*{-1.2cm}\mbox{\!\!\!\includegraphics[width=.12\textwidth]{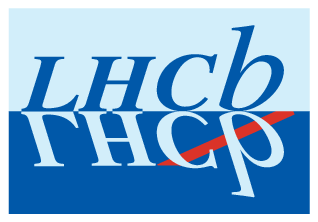}} & &}%
\\
 & & CERN-PH-EP-2014-150 \\  % ID 
 & & LHCb-PAPER-2014-026 \\  % ID 
 & & July 8, 2014 \\ % Date - Can also hardwire e.g.: 23 March 2010
 & & \\
% not in paper \hline
\end{tabular*}

\vspace*{3.cm}

% Title --------------------------------------------------
{\bf\boldmath\huge
\begin{center}
  Measurement of \CP violation in \BsPP decays
\end{center}
}

\vspace*{2.0cm}

% Authors -------------------------------------------------
\begin{center}
The LHCb collaboration\footnote{Authors are listed at the end of the paper.}
\end{center}

\vspace{\fill}

% Abstract -----------------------------------------------
\begin{abstract}
  \noindent
  A measurement of the decay time dependent \CP-violating asymmetry 
  in $\Bs \to \phi\phi$ decays is presented, along with measurements of the \T-odd
  triple-product asymmetries. In this decay channel,
  the \CP-violating weak phase arises from the interference between
  \Bs-\Bsb mixing and the loop-induced decay amplitude. Using a sample of proton-proton collision data 
  corresponding to an integrated luminosity of $3.0\invfb$ collected with the \lhcb detector, 
  a signal yield of approximately 4000 \BsPP decays is obtained. The \CP-violating
  phase is measured to be ${\phisPP=-0.17\pm0.15\stat\pm0.03\syst}\rad$. The triple-product asymmetries are measured to be 
  ${A_U=-0.003\pm0.017\stat\pm0.006\syst}$ and ${A_V=-0.017\pm0.017\stat\pm0.006\syst}$. Results are consistent with the 
  hypothesis of \CP conservation.
\end{abstract}

\vspace*{2.0cm}

\vspace{\fill}

{\footnotesize 
\centerline{\copyright~CERN on behalf of the \lhcb collaboration, license \href{http://creativecommons.org/licenses/by/4.0/}{CC-BY-4.0}.}}
\vspace*{2mm}

\end{titlepage}

%%%%%%%%%%%%%%%%%%%%%%%%%%%%%%%%
%%%%%  EOD OF TITLE PAGE  %%%%%%
%%%%%%%%%%%%%%%%%%%%%%%%%%%%%%%%

%  empty page follows the title page ----
\newpage
\setcounter{page}{2}
\mbox{~}
%\newpage

% Author List ----------------------------
%\input{LHCb_HD_authorlist_2014-05-13.tex}

\cleardoublepage

\renewcommand{\thefootnote}{\arabic{footnote}}
\setcounter{footnote}{0}

%%%%%%%%%%%%%%%%%%%%%%%%%%%%%%%%
%%%%%  Table of Content   %%%%%%
%%%%%%%%%%%%%%%%%%%%%%%%%%%%%%%%
%%%% Uncomment next 2 lines if desired
%\tableofcontents
%\cleardoublepage

%%%%%%%%%%%%%%%%%%%%%%%%%
%%%%% Main text %%%%%%%%%
%%%%%%%%%%%%%%%%%%%%%%%%%

\pagestyle{plain} % restore page numbers for the main text
\setcounter{page}{1}
\pagenumbering{arabic}

\section{Introduction}
\label{sec:Introduction}

The \BsPP  decay is forbidden at tree level in the Standard Model (SM) 
and proceeds predominantly via a gluonic $\bquarkbar \to \squarkbar \ssbar$ loop (penguin) process. 
Hence, this channel provides an excellent probe of new heavy particles entering 
the penguin quantum loops~\cite{Bartsch:2008ps,beneke,PhysRevD.80.114026}.
In the SM, \CP violation is governed by a single phase in the Cabibbo-Kobayashi-Maskawa quark mixing matrix
\cite{Kobayashi:1973fv,*Cabibbo:1963yz}. 
Interference between the \Bs-\Bsb oscillation and decay amplitudes leads to a \CP asymmetry
in the decay time distributions of $\Bs$ and $\Bsb$ mesons, which is characterised by a \CP-violating weak phase. 
Due to different decay amplitudes  the actual value of the weak phase is dependent on the \Bs decay channel.
For  $\Bs \to \jpsi\Kp\Km$ and $\Bs \to \jpsi\pip\pim$ decays, which proceed via $\bquarkbar \to \squarkbar\ccbar$ transitions, the SM prediction of the weak phase
is given by $- 2 \arg \left( -V_{ts} V_{tb}^*/ V_{cs} V_{cb}^*\right)=-0.0364\pm0.0016\; \rm rad$~\cite{Charles:2011va}.
The LHCb collaboration has measured the weak phase in the combination of $\Bs \to \jpsi\Kp\Km$ and $\Bs \to \jpsi\pip\pim$ decays
to be $0.07\pm0.09\stat \pm0.01\syst\rad$~\cite{LHCb-PAPER-2013-002}. A recent analysis
of $\Bs \to \jpsi\pip\pim$ decays using the full \lhcb Run~I dataset of 3.0\invfb has measured the \CP-violating
phase to be $0.070\pm0.068\stat \pm0.008\syst\rad$~\cite{Aaij:2014dka}.
These measurements are consistent with the SM and place stringent constraints on
\CP violation in \Bs-\Bsb oscillations~\cite{LHCb-PAPER-2012-031}. 
The \CP-violating phase, \phisPP, in the \BsPP decay is expected to be small in the SM.
Calculations using quantum chromodynamics factorisation (QCDf) provide an upper limit of 0.02\rad for $|\phisPP|$~\cite{Bartsch:2008ps,beneke,PhysRevD.80.114026}.

Triple-product asymmetries are formed from \T-odd combinations of the momenta of the final state particles. Such asymmetries provide a method of
measuring \CP violation in a decay time integrated method that complements the decay time
dependent measurement~\cite{gronau}. These asymmetries are calculated from functions of the
angular observables and are expected to be close to zero in the SM~\cite{Datta:2012ky}. 
Particle-antiparticle oscillations reduce non-zero triple-product asymmetries due to 
\CP-conserving strong phases, known as ``fake'' triple-product asymmetries by a factor $\Gamma/(\Delta m)$,
where $\Gamma$ and $\Delta m$ are the decay rates and oscillation frequencies of the neutral meson system in question. 
Since one has $\Gs/(\dms)\approx0.04$ for the \Bs system, ``fake'' triple-product asymmetries are strongly suppressed,
allowing for ``true'' \CP-violating triple-product asymmetries to be calculated without
the need to measure the initial flavour of the \Bs meson~\cite{gronau}.

Theoretical calculations can be tested further with measurements of the polarisation fractions, where 
the longitudinal and transverse polarisation fractions are denoted by $f_L$ and $f_T$, respectively.
In the heavy quark limit, $f_L$ is expected to be the dominant polarisation
due to the vector-axial structure of charged weak currents~\cite{beneke}. 
This is found to be the case for tree-level \PB decays measured at the 
\PB factories~\cite{PhysRevLett.94.221804,PhysRevLett.98.051801,PhysRevD.78.092008, delAmoSanchez:2010mz,Abe:2004mq, Aubert:2006fs}.
However, the dynamics of penguin transitions are more complicated. In the context
of QCDf, $f_L$ is predicted to be $0.36^{+0.23}_{-0.18}$ for the \BsPP decay~\cite{PhysRevD.80.114026}.

In this paper, a measurement of the \CP-violating phase in $\Bs\to \phi(\to\Kp\Km) \phi(\to\Kp\Km)$ decays, along
with a measurement of the \T-odd triple-product asymmetries is presented. 
The results are based on $pp$ collision data
corresponding to an integrated luminosity of  $1.0\invfb$ and $2.0\invfb$ 
collected by the \lhcb experiment at centre-of-mass energies $\sqrt{s}=7\tev$ in 2011 and 8\tev in 2012, respectively. 
Previous measurements of the triple-product asymmetries from the \lhcb~\cite{LHCb-PAPER-2012-004} and \cdf~\cite{Aaltonen:2011rs} collaborations,
together with the first measurement of the \CP-violating phase in \BsPP decays~\cite{LHCb-PAPER-2013-007}, have shown no evidence
of deviations from the SM. The decay time dependent measurement improves on the previous analysis~\cite{LHCb-PAPER-2013-007}
through the use of a more efficient candidate selection and improved knowledge of the \Bs flavour
at production, in addition to a data-driven determination of the efficiency as a function of decay time.

The results presented in this paper supersede previous measurements of the \CP-violating phase~\cite{LHCb-PAPER-2013-007}
and \T-odd triple-product asymmetries~\cite{LHCb-PAPER-2012-004}, made using 
1.0\invfb of data collected at a $\sqrt{s}=7\tev$.

\section{Detector description}
\label{sec:Detector}

The \lhcb detector~\cite{Alves:2008zz} is a single-arm forward
spectrometer covering the \mbox{pseudorapidity} range $2<\eta <5$,
designed for the study of particles containing \bquark or \cquark
quarks. The detector includes a high-precision tracking system
consisting of a silicon-strip vertex detector surrounding the $pp$
interaction region, a large-area silicon-strip detector located
upstream of a dipole magnet with a bending power of about
$4{\rm\,Tm}$, and three stations of silicon-strip detectors and straw
drift tubes~\cite{LHCb-DP-2013-003} placed downstream.
The combined tracking system provides a momentum measurement with
relative uncertainty that varies from 0.4\% at low momentum to 0.6\% at 100\gevc,
and impact parameter resolution of 20\mum for
tracks with large transverse momentum, \pt. Different types of charged hadrons are distinguished using information
from two ring-imaging Cherenkov (RICH) detectors~\cite{LHCb-DP-2012-003}. Photon, electron and
hadron candidates are identified by a calorimeter system consisting of
scintillating-pad and preshower detectors, an electromagnetic
calorimeter and a hadronic calorimeter.
The trigger~\cite{LHCb-DP-2012-004} consists of a
hardware stage, based on information from the calorimeter and muon
systems, followed by a software stage, which applies a full event
reconstruction.
The hardware trigger selects \BsPP candidates by requiring large transverse
energy deposits in the calorimeters from at least one of the 
final state particles.
In the software trigger, \BsPP candidates are selected 
either by identifying events containing a pair of oppositely charged kaons 
with an invariant mass close to that of the  $\phi$ meson or by using a topological \bquark-hadron trigger.
The topological software trigger requires a two-, three- or four-track
secondary vertex with a large sum of the \pt of
the charged particles and a significant displacement from the primary $pp$
interaction vertices~(PVs). At least one charged particle should have $\pt >
1.7\gevc$ and \chisqip with respect to any
primary interaction greater than 16, where \chisqip is defined as the
difference in \chisq of a given PV fitted with and
without the considered track.
A multivariate algorithm~\cite{BBDT} is used for
the identification of secondary vertices consistent with the decay
of a \bquark-hadron.

In the simulation, $pp$ collisions are generated using
\pythia~\cite{Sjostrand:2006za,*Sjostrand:2007gs} with a specific \lhcb
configuration~\cite{LHCb-PROC-2010-056}.  Decays of hadronic particles
are described by \evtgen~\cite{Lange:2001uf}, in which final state
radiation is generated using \photos~\cite{Golonka:2005pn}. The
interaction of the generated particles with the detector and its
response are implemented using the \geant
toolkit~\cite{Allison:2006ve, *Agostinelli:2002hh} as described in
Ref.~\cite{LHCb-PROC-2011-006}.

\section{Selection and mass model}
\label{sec:selection}

Events passing the trigger are initially required to
pass loose requirements on the fit quality of the four-kaon vertex fit, the \chisqip of each
track, the transverse momentum of each particle, and the product of the transverse 
momenta of the two $\phi$ candidates. In addition, the reconstructed mass of 
$\phi$ meson candidates is required to be within 25\mevcc of the known $\phi$ mass~\cite{PDG2012}.

In order to further separate the \BsPP signal from the background, a boosted decision tree (BDT)
is implemented~\cite{Breiman,AdaBoost}. To train the BDT, simulated
\BsPP events passing the same loose requirements as the data events are used as signal,
whereas events in the four-kaon invariant mass sidebands from data are used as background. The signal mass region is
defined to be less than 120\mevcc from the known \Bs mass, $m_{\Bs}$~\cite{PDG2012}. The invariant mass
sidebands are defined to be inside the region $120<|m_{\Kp\Km\Kp\Km}-m_{\Bs}|<300\mevcc$, where $m_{\Kp\Km\Kp\Km}$ is the four-kaon invariant mass.
Separate BDTs are trained for data samples collected in 2011 and 2012, due to different data taking conditions
in the different years.
Variables used in the BDT consist of the minimum and maximum kaon \pt and $\eta$, the minimum and the maximum
\pt and $\eta$ of the $\phi$ candidates, the \pt and $\eta$ of the \Bs candidate, the minimum probability of the kaon mass hypothesis using information
from the RICH detectors, the quality of the four-kaon vertex fit, and
the \chisqip of the \Bs candidate.
The BDT also includes kaon isolation asymmetries. The isolation variable is 
calculated as the scalar sum of the \pt of charged particles inside a region defined as
$\sqrt{\Delta\varphi^2+\Delta\eta^2}<1$, where $\Delta\varphi\,(\Delta\eta)$ is the difference
in azimuthal angle (pseudorapidity), not including the signal kaon from the \Bs decay.
The asymmetry is then calculated as the difference between the isolation variable and the \pt
of the signal kaon, divided by the sum.
After the BDT is trained, the optimum requirement on each BDT is chosen to maximise $N_{\rm S}/\sqrt{N_{\rm S}+N_{\rm B}}$, where
${N_{\rm S}\,(N_{\rm B})}$ represent the expected number of signal (background) events in the signal region
of the data sample.

The presence of peaking backgrounds is extensively studied. The decay modes considered include $\Bp\to\phi\Kp$, $\Bd\to\phi\pip\pim$, $\Bd\to\phi\Kstarz$, 
and $\Lb\to\phi \proton\Km$, of which only the last two are found to contribute, and are the result of a mis-identification
of a pion or proton as a kaon, respectively. 
The number of $\Bd\to\phi\Kstarz$ events present in the data sample is determined from 
scaling the number of $\Bd\to\phi\Kstarz$ events seen in data through a different dedicated selection with the relative
efficiencies between the two selections found from simulated events.
This method yields values of $7.3\pm0.4$ and $17.8\pm0.9$ events in the 2011 and 2012 datasets, respectively.
The amount of $\Lb\to\phi \proton\Km$ decays is estimated directly from
data by changing the mass hypothesis of the final-state particle most likely to have the mass of the proton
from RICH detector information. This method yields $52\pm19$ and $51\pm 29$ $\Lb\to\phi \proton\Km$ events
in the 2011 and 2012 datasets, respectively. 

\begin{figure}[t]
\centering
\includegraphics[width=0.49\textwidth]{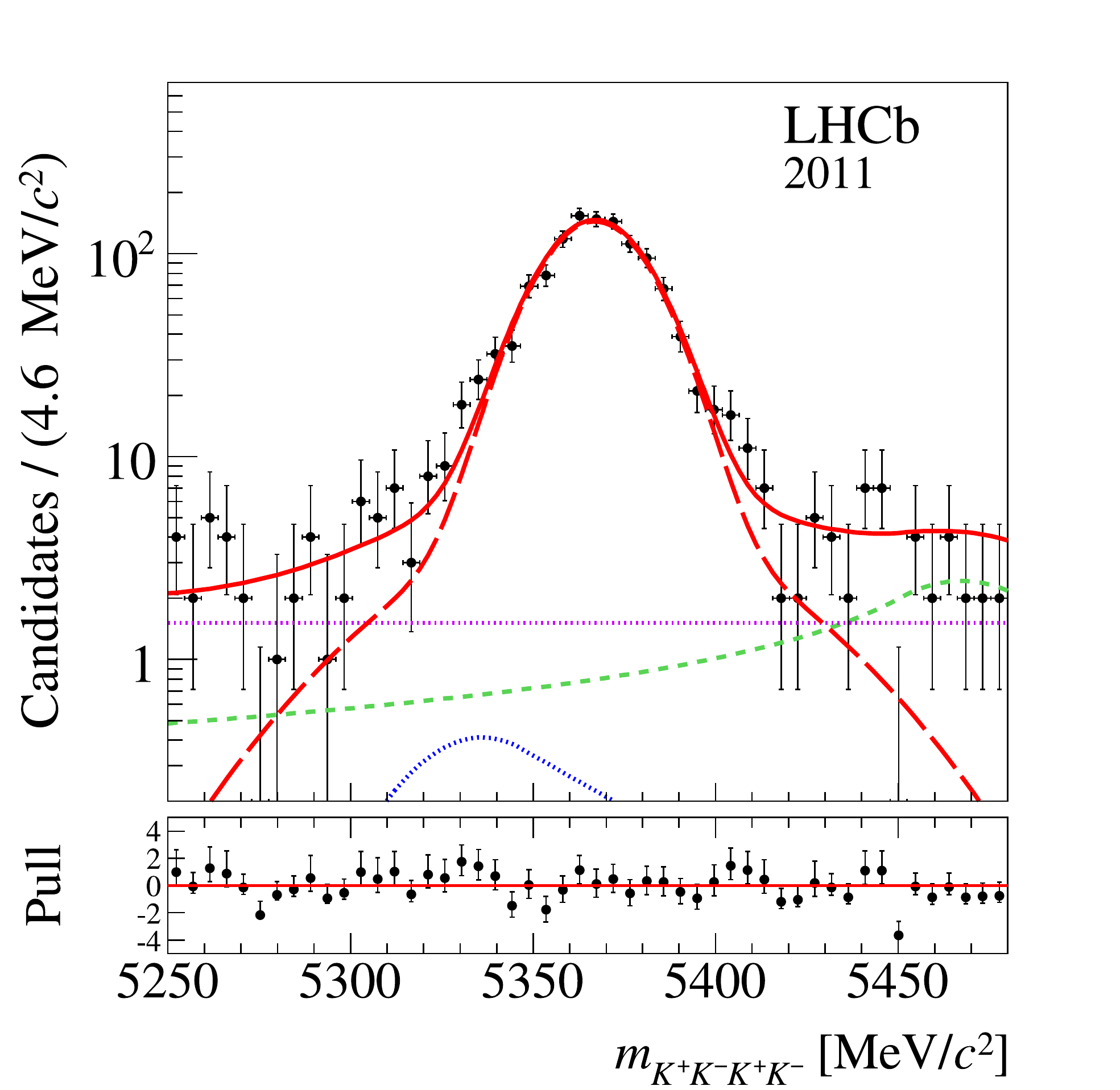}
\includegraphics[width=0.49\textwidth]{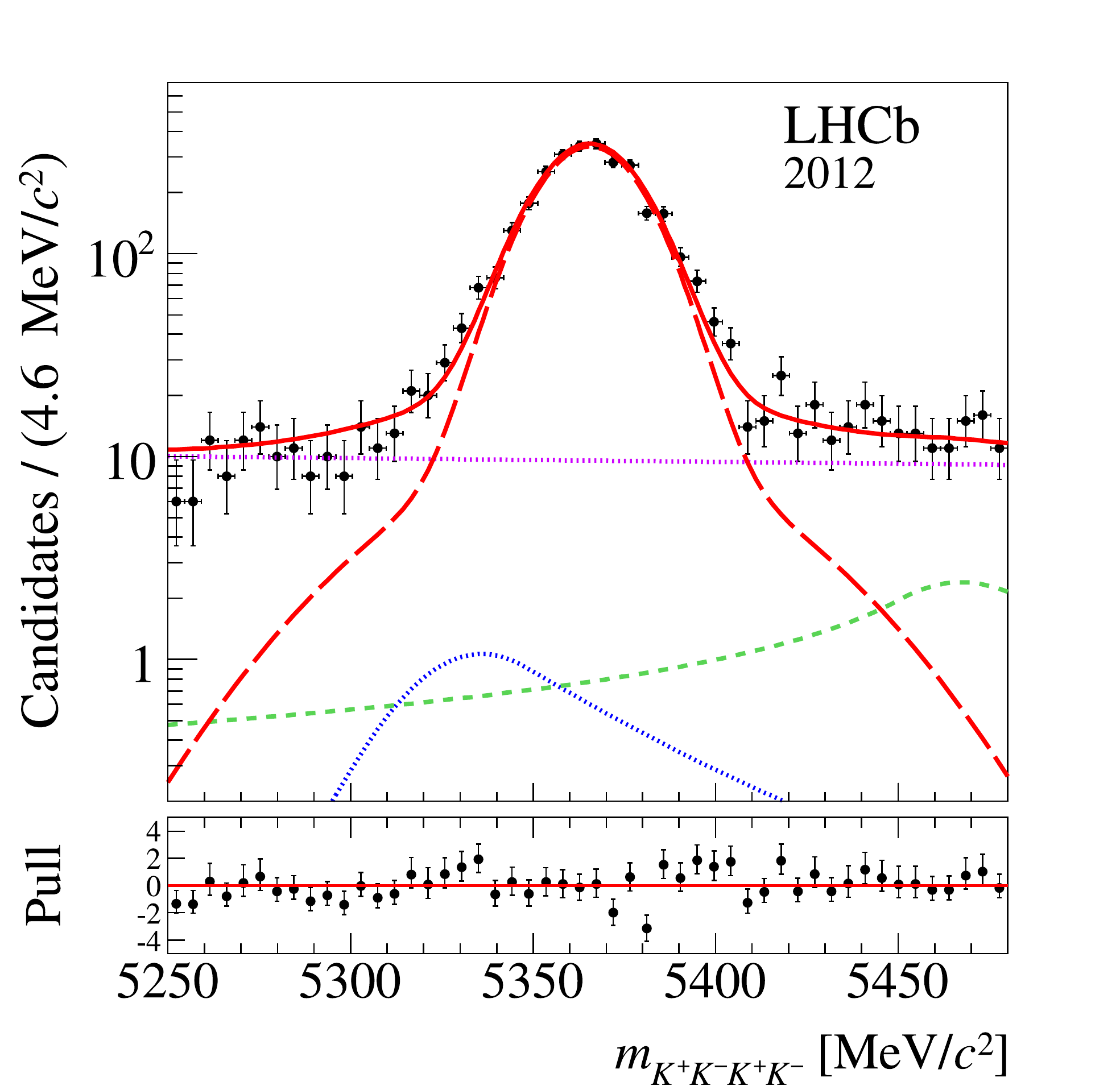}
\caption{\small Four-kaon invariant mass distributions for the (left) 2011 and (right) 2012 datasets.
The data points are represented by the black markers. Superimposed are the results of the total fit 
(red solid line), the \BsPP (red long dashed), the $\Bd\to\phi\Kstarz$ (blue dotted),
the $\Lb\to\phi \proton\Km$ (green short-dashed), and the combinatoric (purple dotted) fit components.
}
\label{fig:massPlots}
\end{figure}

In order to correctly determine the number of \BsPP events in the final data sample, 
the four-kaon invariant mass distributions are fitted with
the \BsPP signal described by a double Gaussian model, and the combinatorial background component described using
an exponential function. 
The peaking background contributions are fixed to the shapes found in simulated events. The yields of the peaking
background contributions are fixed to the numbers previously stated.
This consists of the sum of a Crystal Ball function~\cite{Skwarnicki:1986xj} and a Gaussian to describe the $\Bd\to\phi\Kstarz$ reflection
and a Crystal Ball function to describe the $\Lb\to\phi \proton\Km$ reflection.
Once the BDT requirements are imposed, an unbinned extended maximum likelihood 
fit to the four-kaon invariant mass yields $1185\pm35$ and $2765 \pm 57$ \BsPP events in the 2011 and 2012 datasets,
respectively. The combinatorial background yield is found to be $76\pm17$ and $477\pm32$ in the 2011 and 2012 datasets,
respectively. The fits to the four-kaon invariant mass are shown in Fig.~\ref{fig:massPlots}.

The use of the four-kaon invariant mass to assign signal weights
allows for a decay time dependent fit to be performed with
only the signal distribution explicitly described.
The method for assigning the signal weights is described in greater detail in Sec.~\ref{sec:LL_TD}.

\section{Phenomenology}
\label{sec:phenom}

\begin{figure}[ht]
\centering
\includegraphics[width=140mm]{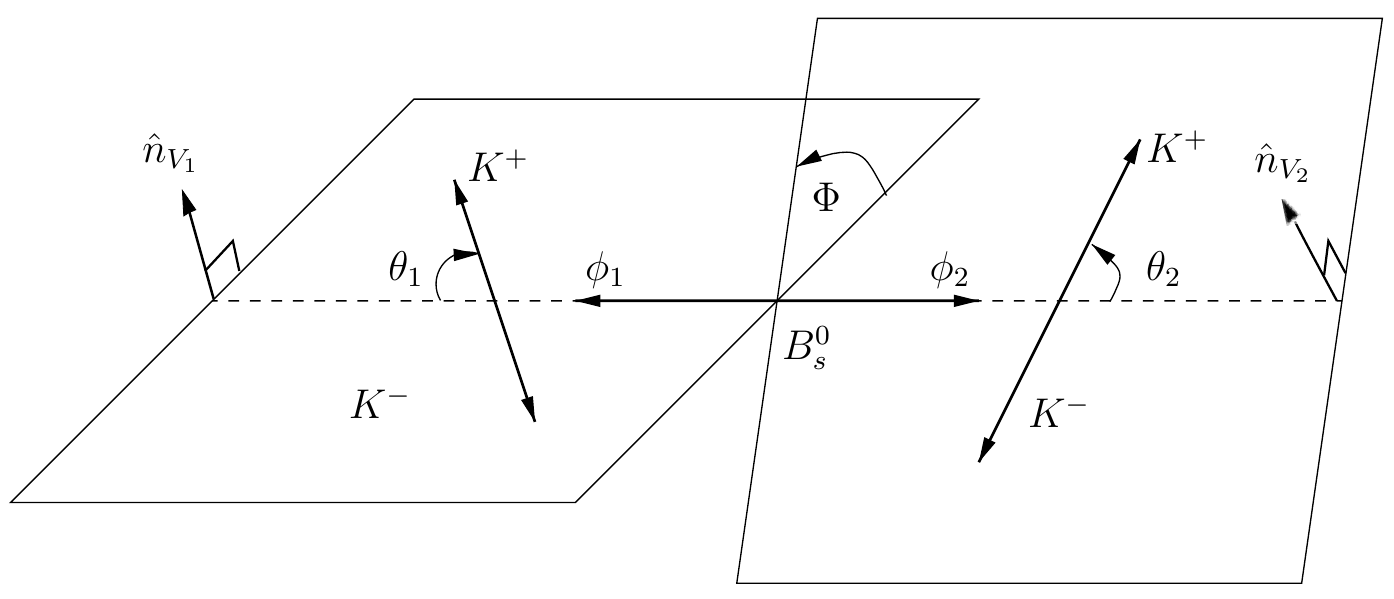}
\caption{\small Decay angles for the $\Bs \rightarrow \phi \phi$ decay, where the $K^+$ momentum in the $\phi_{1,2}$ rest frame
and the parent $\phi_{1,2}$ momentum in the rest frame of the \Bs meson span the two $\phi$ meson decay planes,
$\theta_{1,2}$ is the angle between the $K^+$ track momentum in the $\phi_{1,2}$ meson rest frame and
the parent $\phi_{1,2}$ momentum in the \Bs rest frame, $\Phi$ is the angle between the two $\phi$ meson
decay planes and $\hat{n}_{V_{1,2}}$ is the unit vector normal to the decay plane of the $\phi_{1,2}$ meson.}
\label{fig:angles}
\end{figure}

The \BsPP decay is composed of a mixture of \CP eigenstates,
that are disentangled by means of an angular analysis
in the helicity basis, defined in Fig.~\ref{fig:angles}.

\subsection{Decay time dependent model}
\label{sec:modelTD}

The \BsPP decay is a  $P\to VV$ decay, where $P$ denotes a pseudoscalar and $V$ a vector meson.
However, due to the proximity of the $\phi$ 
resonance to that of the $f_0(980)$, 
there will also be contributions from $S$-wave ($P\to V\kern-0.1em S$) and double $S$-wave ($P\to S\kern-0.1em S$) processes, where $S$ denotes 
a spin-0 meson or a pair of non-resonant kaons.
Thus the total amplitude is a coherent sum of $P$-, $S$-, and double $S$-wave processes, and is
accounted for during fitting
by making use of the different functions of the helicity angles associated with these terms.
The choice of which $\phi$ meson is used to determine $\theta_1$ and which is
used to determine $\theta_2$ is randomised.
The total amplitude ($\mathcal{A}$) containing the $P$-, $S$-, and double $S$-wave components as
a function of decay time, $t$, can be written as~\cite{Bhattacharya:2013sga}
\begin{align}
\mathcal{A}(t,\theta_1,\theta_2,\Phi) &= A_0(t)\cos\theta_1\cos\theta_2 
+\frac{A_\parallel(t)}{\sqrt{2}}\sin\theta_1\sin\theta_2\cos\Phi \nonumber \\
&+i\frac{A_\perp(t)}{\sqrt{2}}\sin\theta_1\sin\theta_2\sin\Phi
+\frac{A_S(t)}{\sqrt{3}}(\cos\theta_1 + \cos\theta_2)
+\frac{A_{SS}(t)}{3},
\end{align}
where $A_0$, $A_\parallel$, and $A_\perp$ are the \CP-even longitudinal, \CP-even parallel, and \CP-odd perpendicular
polarisations of the \BsPP decay. The $P\to V\kern-0.1em S$ and $P\to S\kern-0.1em S$ processes are described by the $A_S$ and $A_{SS}$ amplitudes, respectively.
The differential decay rate may be found through the square of the total amplitude leading to 
the 15 terms~\cite{Bhattacharya:2013sga}
\begin{align}
\frac{\deriv \Gamma}{\deriv t\,\deriv \cos\theta_1\,\deriv \cos\theta_2 \,\deriv\Phi} \propto 4|\mathcal{A}(t,\theta_1,\theta_2,\Phi)|^2
= \sum_{i=1}^{15} K_i (t) f_i (\theta_1,\theta_2,\Phi).
\label{eq:pdf}
\end{align}
The $K_i(t)$
term can be written as
\begin{align}
K_i(t)=N_ie^{-\Gs t} \left[ c_i \cos(\dms t) + d_i\sin(\dms t)
+ a_i \cosh\left(\frac{1}{2}\DGs t\right) + b_i \sinh\left(\frac{1}{2}\DGs t\right) \right]\kern-0.25em ,
\label{eq:pdfK}
\end{align}
where the coefficients are shown in Table~\ref{tab:terms},
$\DGs \equiv \Gamma_{\rm L} - \Gamma_{\rm H}$ is the decay width difference between the
light and heavy \Bs mass eigenstates,
 $\Gs \equiv (\Gamma_{\rm L }+ \Gamma_{\rm H})/2$ is the average decay width, 
  and $\dms$ is the \Bs-\Bsb oscillation frequency. 
The differential decay rate for a \Bsb meson produced at $t=0$ is obtained
by changing the sign of the $c_i$ and $d_i$ coefficients.
\begin{table}[t]
\begin{center}
\caption{\label{tab:terms} \small Coefficients of the time dependent terms and angular functions used in Eq.~\ref{eq:pdf}. Amplitudes are defined at $t=0$.}
{
\scriptsize
$
\arraycolsep=1.6pt\def\arraystretch{1.4}
%\[
\begin{array}{c|c|c|c|c|c|c}
i     & N_i                                 & a_i         & b_i                         & c_i                         & d_i          & f_i \\ \hline
1       & |A_0|^2                          & 1             & D                  & C                             & -S   & 4\cos^2\theta_1\cos^2\theta_2 \\
2       & |A_\parallel |^2                 & 1             & D                  & C                             & -S   & \sin^2\theta_1\sin^2\theta_2(1{+}\cos 2\Phi) \\
3       & |A_\perp |^2                     & 1             & -D                   & C                             & S  & \sin^2\theta_1\sin^2\theta_2(1{-}\cos 2\Phi) \\
4       & |A_\parallel||A_\perp |       & C\sin\delta_1             & S\cos\delta_1      & \sin\delta_1      & D\cos\delta_1 & -2\sin^2\theta_1\sin^2\theta_2\sin 2\Phi \\
5       & |A_\parallel||A_0  |          & \cos(\delta_{2,1})             & D\cos(\delta_{2,1})                   & C\cos\delta_{2,1}                             &  -S\cos(\delta_{2,1})   & \sqrt{2}\sin2\theta_1\sin2\theta_2\cos\Phi \\
6       & |A_0 ||A_\perp|               & C\sin\delta_2             & S\cos\delta_2      & \sin\delta_2      & D\cos\delta_2& -\sqrt{2}\sin2\theta_1\sin2\theta_2\sin\Phi \\
7       & |A_{SS}|^2                       & 1             & D                  & C                             & -S   & \frac{4}{9} \\
8       & |A_S|^2                          & 1             & -D                   & C                             & S  & \frac{4}{3}(\cos\theta_1+\cos\theta_2)^2 \\
9       & |A_S| |A_{SS} |               & C\cos(\delta_S-\delta_{SS})             & S\sin(\delta_S{-}\delta_{SS}) & \cos(\delta_{SS}{-}\delta_{S}) & D\sin(\delta_{SS}{-}\delta_{S}) & \frac{8}{3\sqrt{3}}(\cos\theta_1+\cos\theta_2)\\
10      & |A_0 | |A_{SS}| & \cos\delta_{SS}            & D\cos\delta_{SS}                  & C\cos\delta_{SS}                             & -S\cos\delta_{SS}   & \frac{8}{3}\cos\theta_1\cos\theta_2 \\
11      & |A_\parallel | |A_{SS} | &  \cos(\delta_{2,1}{-}\delta_{SS})               & D\cos(\delta_{2,1}{-}\delta_{SS})                  & C\cos(\delta_{2,1}{-}\delta_{SS})             &  -S\cos(\delta_{2,1}{-}\delta_{SS})   & \frac{4\sqrt{2}}{3}\sin\theta_1\sin\theta_2\cos\Phi \\
12      & |A_\perp | |A_{SS} |          & C\sin(\delta_2-\delta_{SS})             & S\cos(\delta_2-\delta_{SS})& \sin(\delta_2{-}\delta_{SS}) & D\cos(\delta_2{-}\delta_{SS}) &  -\frac{4\sqrt{2}}{3}\sin\theta_1\sin\theta_2\sin\Phi \\[1ex]
\multirow{2}{*}{$13$}      &\multirow{2}{*}{$ |A_0 | |A_{S} | $}               &\multirow{2}{*}{$C\cos\delta_S$}             &\multirow{2}{*}{$ -S\sin\delta_S$}                & \multirow{2}{*}{$\cos\delta_S $}               & \multirow{2}{*}{ $ -D\sin\delta_S$} 
& \frac{8}{\sqrt{3}} \cos\theta_1\cos\theta_2\\
& & & & & & \times (\cos\theta_1 + \cos\theta_2) \\[1.5ex]
\multirow{2}{*}{$14$}      &\multirow{2}{*}{$ |A_\parallel | |A_{S} | $}      & \multirow{2}{*}{$C\cos(\delta_{2,1}-\delta_S)$}             & \multirow{2}{*}{$ S\sin(\delta_{2,1}-\delta_S) $} &\multirow{2}{*}{$ \cos(\delta_{2,1}{-}\delta_S)  $}     &\multirow{2}{*}{$  D\sin(\delta_{2,1}-\delta_S)$} 
& \frac{4\sqrt{2}}{\sqrt{3}} \sin\theta_1\sin\theta_2 \\
& & & & & & \times (\cos\theta_1 + \cos\theta_2)\cos\Phi \\[1.5ex]
\multirow{2}{*}{$15$}      & \multirow{2}{*}{$ |A_\perp | |A_{S}| $} &  \multirow{2}{*}{$ \sin(\delta_2-\delta_S) $}  & \multirow{2}{*}{$-D \sin(\delta_2-\delta_S) $}                  & \multirow{2}{*}{$C\sin(\delta_2-\delta_S)$}                             & \multirow{2}{*}{$S\sin(\delta_2-\delta_S)$} 
& -\frac{4\sqrt{2}}{\sqrt{3}} \sin\theta_1\sin\theta_2 \\
& & & & & & \times (\cos\theta_1 + \cos\theta_2)\sin\Phi \\[1.5ex]
\end{array}
$
}
\end{center}
\end{table}

The three \CP-violating terms introduced in Table~\ref{tab:terms} are defined as
\begin{align}
C &\equiv \frac{1-|\lambda|^2}{1+|\lambda|^2}, \\
S &\equiv -\frac{2|\lambda|\sin\phisPP}{1+|\lambda|^2},\\
D &\equiv -\frac{2|\lambda|\cos\phisPP}{1+|\lambda|^2},
\end{align}
where \phisPP measures \CP violation in the interference between the direct
decay amplitude and that via mixing, $\lambda \equiv (q/p)(\overline{A}/A)$,
$q$ and $p$ are the complex
parameters relating the \Bs flavour and mass eigenstates, and $A\,(\overline{A})$ is the decay amplitude (\CP conjugate decay amplitude).
Under the assumption that $|q/p|=1$, $|\lambda|$ measures direct \CP violation.
The \CP violation parameters are assumed to be helicity independent. 
The association of \phisPP and $|\lambda|$ with $S$-wave and double $S$-wave terms implies that these consist solely of contributions
with the same flavour content as the $\phi$ meson, i.e. an \ssbar resonance.

In Table~\ref{tab:terms},
$\delta_S$ and $\delta_{SS}$ are the strong phases of the $P\to V\kern-0.1em S$ and $P\to S\kern-0.1em S$ processes, respectively. 
The $P$-wave strong phases are defined to be $\delta_1\equiv\delta_\perp-\delta_\parallel$ and $\delta_2\equiv\delta_\perp-\delta_0$,
with the notation $\delta_{2,1}\equiv\delta_2-\delta_1$.

\subsection{Triple-product asymmetries}
\label{sec:modelTP}

Scalar triple products of three momentum or spin vectors are odd under
time reversal, \T. Non-zero asymmetries for these observables can either be due to a \CP-violating phase or a
\CP-conserving phase and final-state interactions. 
Four-body final states give rise to three independent momentum vectors in the rest frame of the decaying \Bs
meson. For a detailed review of the phenomenology the reader is referred to Ref.~\cite{gronau}. 

The two independent terms in the time dependent decay rate that contribute to a $T$-odd asymmetry 
are the $K_4(t)$ and $K_6(t)$ terms, defined in Eq.~\ref{eq:pdfK}. The triple products that allow access to these terms are 
\begin{eqnarray}
\sin \Phi = (\hat{n}_{V_1} \times \hat{n}_{V_2}) \cdot \hat{p}_{V_1}, \\
\sin 2\Phi = 2(\hat{n}_{V_1} \cdot \hat{n}_{V_2})(\hat{n}_{V_1} \times \hat{n}_{V_2}) \cdot \hat{p}_{V_1},
\end{eqnarray}
where $\hat{n}_{V_i}$ ($i = 1,2$) is a unit vector perpendicular to the $V_i$ decay 
plane and $\hat{p}_{V_1}$ is a unit vector in the direction of $V_1$ in the \Bs rest frame, defined in
Fig.~\ref{fig:angles}.
This then provides a method of probing \CP violation without the need to measure the decay time 
or the initial flavour of the \Bs meson.
It should be noted that while the observation of non-zero triple-product asymmetries implies \CP violation or final state interactions
(in the case of \Bs meson decays), the measurements of triple-product asymmetries consistent with zero
do not rule out the presence of \CP-violating effects, as strong phase differences can cause suppression~\cite{gronau}.

In the $\BsPP$ decay, two triple products are defined as $U \equiv \sin\Phi\cos\Phi$
and $V \equiv \sin(\pm \Phi)$ where the positive sign is taken 
if $\cos \theta_1 \cos \theta_2 \geq 0$ and negative sign
otherwise. 

The \T-odd asymmetry corresponding to the $U$ observable, $A_U$, is defined as the normalised difference between the number of decays
with positive and negative values of $\sin\Phi\cos\Phi$,
\begin{align}
A_U \equiv \frac{\Gamma(U > 0) - \Gamma(U < 0)}{\Gamma(U > 0) + \Gamma(U < 0)} \propto \int^\infty_0 \Im \left( A_\perp(t) A_\parallel^*(t) + \bar{A}_\perp(t) \bar{A}_\parallel^*(t) \right) \deriv t.
\end{align}
Similarly $A_V$ is defined as
\begin{eqnarray}
A_V \equiv \frac{\Gamma(V > 0) - \Gamma(V < 0)}{\Gamma(V > 0) + \Gamma(V < 0)} \propto \int^\infty_0 {\Im} \left( A_\perp(t) A_0^*(t) + \bar{A}_\perp(t) \bar{A}_0^*(t)\right) \deriv t.
\end{eqnarray}

Extraction of the triple-product asymmetries is then reduced to a
simple counting exercise.

\section{Decay time resolution}
\label{sec:DTR}

The sensitivity to \phisPP is affected by the accuracy of the measured decay time.
In order to resolve the fast \Bs-\Bsb oscillation period of approximately $355\fs$, it is necessary to have a decay time 
resolution that is much smaller than this.
To account for decay time resolution, all decay time dependent terms 
are convolved with a Gaussian function, with width $\sigma^t_i$ that
is estimated for each event, $i$, based upon the uncertainty obtained from the vertex and kinematic fit.
In order to apply an event-dependent resolution model during fitting, the estimated per-event decay 
time uncertainty must be calibrated. 
This is done using simulated events that are divided into bins of $\sigma^t_i$. 
For each bin, a Gaussian function is fitted to the difference between reconstructed decay time and the 
true decay time to determine the resolution $\sigma_{\rm true}^t$. 
A first-order polynomial is then fitted to the distribution of $\sigma^t_i$ versus $\sigma_{\rm true}^t$, 
with parameters denoted by $q_0$ and $q_1$.
The calibrated per-event decay time uncertainty used in the decay time dependent fit
is then calculated as $\sigma^{\rm cal}_i = q_0 + q_1 \sigma^t_i$.
Gaussian constraints are used to account for the uncertainties on the calibration parameters
in the decay time dependent fit.
Cross-checks, consisting of the variation of an effective single Gaussian resolution far beyond the observed differences
in data and simulated events yield 
negligible modifications to results, hence no systematic uncertainty is assigned.
The results are verified to be largely insensitive to the details of the resolution model, as supported by tests 
on data and observed in similar measurements~\cite{LHCb-PAPER-2013-002}.

The effective single Gaussian resolution is found from simulated datasets to be
{$41.4\pm 0.5$\fs} and {$43.9\pm0.5$\fs} for the 2011 and 2012 datasets, respectively.
Differences in the resolutions from 2011 and 2012 datasets are expected due to the 
independent selection requirements.

\section{Acceptances}
\label{sec:acc}

The four observables used to analyse \BsPP events consist of the decay time and the three helicity
angles, which require a good understanding of efficiencies in these variables.
It is assumed that the decay time and angular acceptances factorise.

\subsection{Angular acceptance}
\label{sec:AA}

\begin{figure}[t]
\centering
\includegraphics[width=0.45\textwidth]{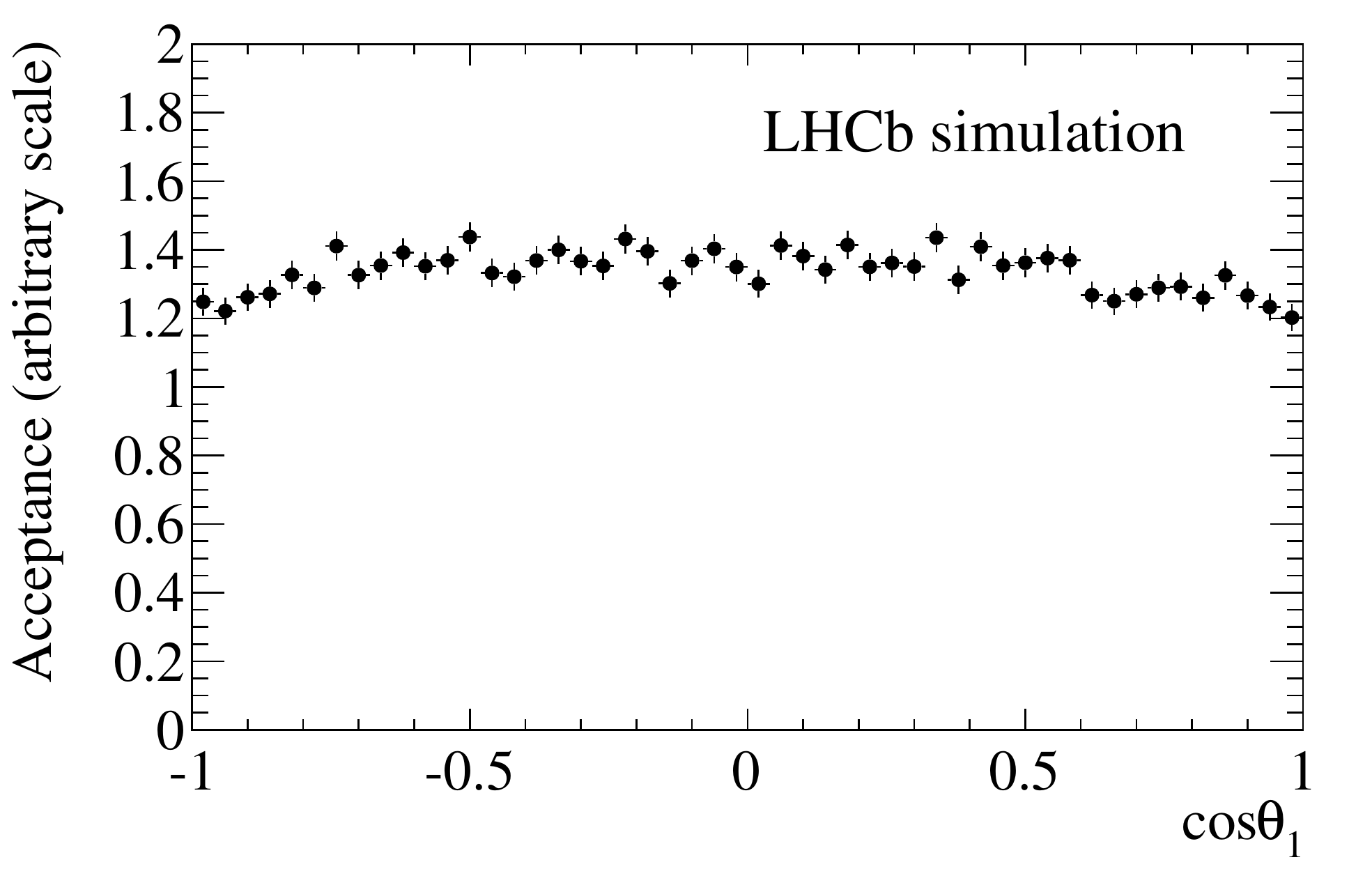}
\includegraphics[width=0.45\textwidth]{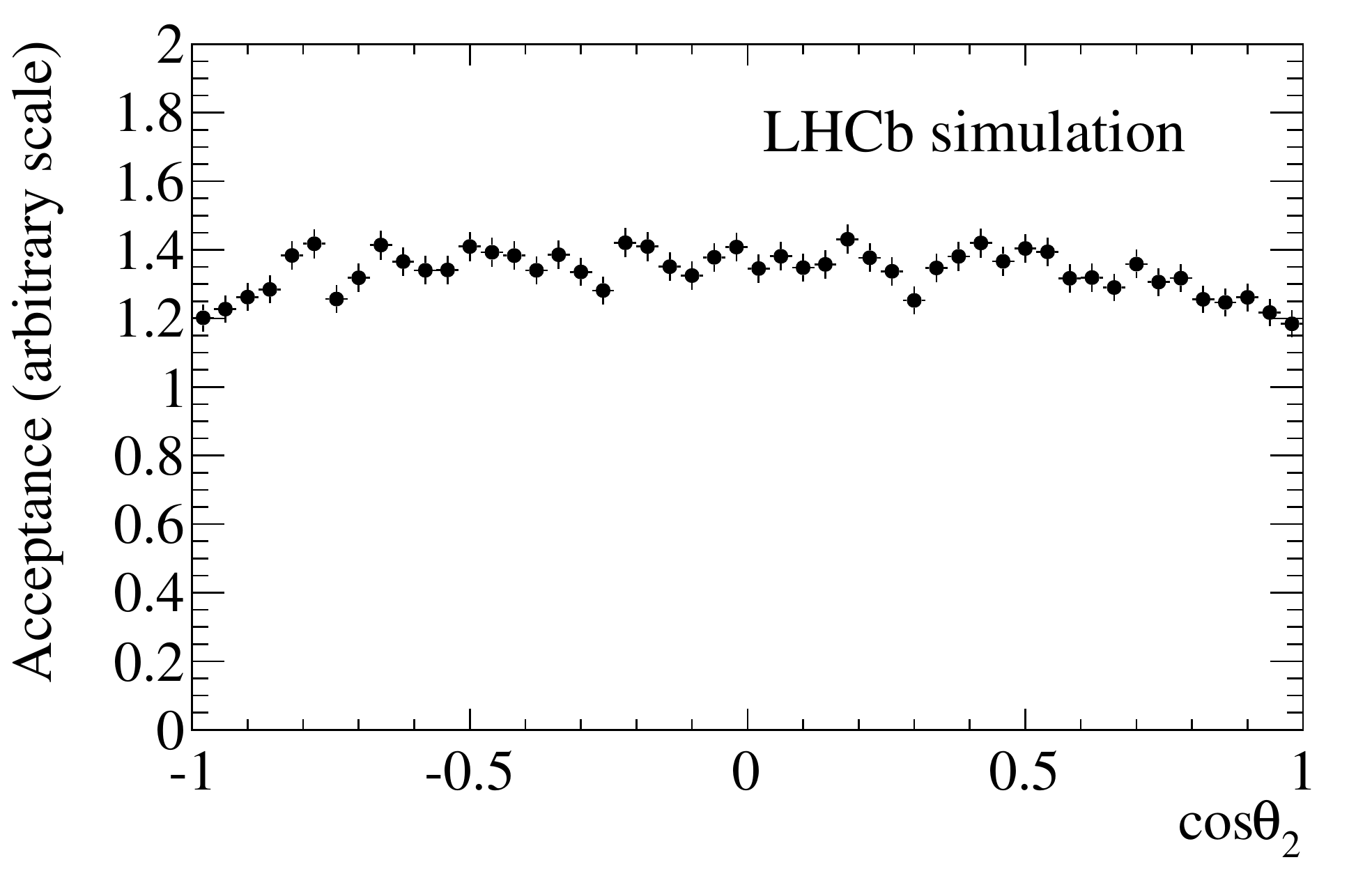}
\includegraphics[width=0.45\textwidth]{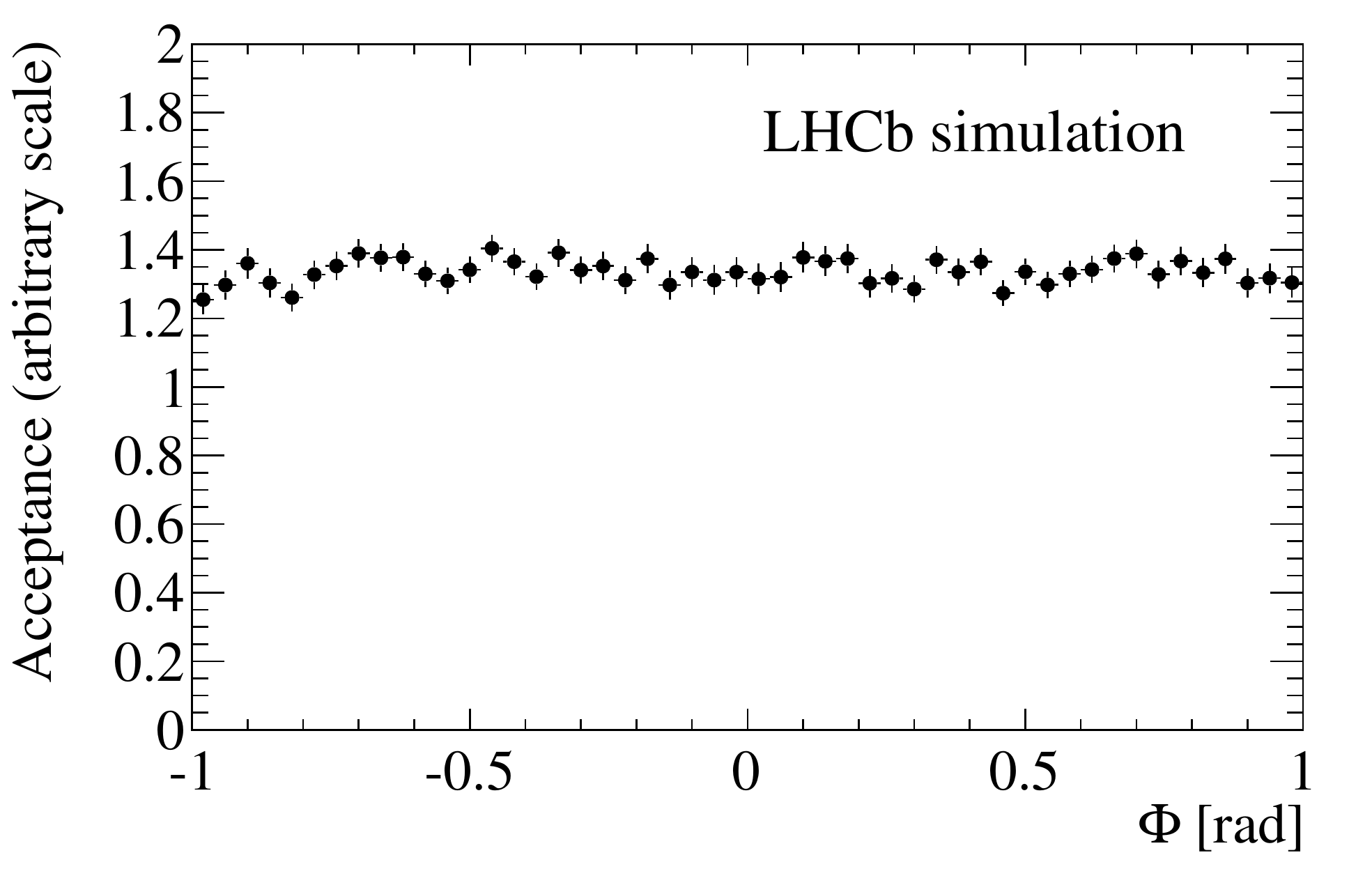}
\caption{
\small Angular acceptance found from simulated \BsPP events (top-left) integrated over $\cos\theta_2$ and $\Phi$ as a function of $\cos\theta_1$, (top-right) integrated over 
$\cos\theta_1$ and $\Phi$ as a function of $\cos\theta_2$, and (bottom) integrated over
$\cos\theta_1$ and $\cos\theta_2$ as a function of $\Phi$.}
\label{fig:AngAcc}
\end{figure}

The geometry of the \lhcb detector and the momentum requirements imposed on the final-state particles
introduce efficiencies that vary as functions of the helicity angles. Simulated events with the 
same selection criteria as those applied to \BsPP data events are used to determine this efficiency correction.
Efficiencies as a function of the three helicity angles are shown in Fig.~\ref{fig:AngAcc}.

Acceptance functions are included in the decay time dependent fit through the 15
integrals $\int\epsilon(\Omega)f_k(\Omega)\deriv\Omega$, where $f_k$ are the angular functions
given in Table~\ref{tab:terms} and $\epsilon(\Omega)$ is the efficiency as a function of the
set of helicity angles, $\Omega$. The inclusion of the integrals in the normalisation of the probability
density function (PDF)
is sufficient to describe the angular acceptance as the acceptance factors for each event
appear as a constant in the log-likelihood, the construction of which is described in detail in Sec.~\ref{sec:LL_TD}, 
and therefore do not affect the fitted parameters.
The method for the calculation of the integrals is described in detail in Ref.~\cite{pree}.
The integrals are calculated correcting for the differences between data and simulated events.
This includes differences in the BDT training variables that can affect acceptance 
corrections through correlations with the helicity angles.

The fit to determine the triple-product asymmetries assumes that the $U$ and $V$ observables
are symmetric in the acceptance corrections. Simulated events are then used to assign a 
systematic uncertainty related to this assumption.

\subsection{Decay time acceptance}
\label{sec:DTA}

The impact parameter requirements on the final-state particles efficiently suppress the background from
numerous pions and kaons originating from the PV, but introduce a decay time dependence in the 
selection efficiency. 

The efficiency as a function of the decay time is taken from $\Bs\to\Dsm(\to \Kp\Km\pim) \pip$ data events,
with an upper limit of 1\ps applied to the \Dsm decay time to ensure topological similarity
to the \BsPP decay.
After the same decay time-biasing selections are applied to the $\Bs\to\Dsm\pip$ decay as used
in the \BsPP decay, $\Bs\to\Dsm\pip$ events are re-weighted according to the minimum track
transverse momentum to ensure the closest agreement between the time acceptances
of \BsPP and $\Bs\to\Dsm\pip$ simulated events.
The denominator used to calculate the decay time acceptance in $\Bs\to\Dsm\pip$ data
is taken from a simulated dataset, generated with the \Bs lifetime taken from the value measured by the
\lhcb experiment~\cite{DspiLT}.

\begin{figure}[t]
\centering
\includegraphics[width=0.45\textwidth]{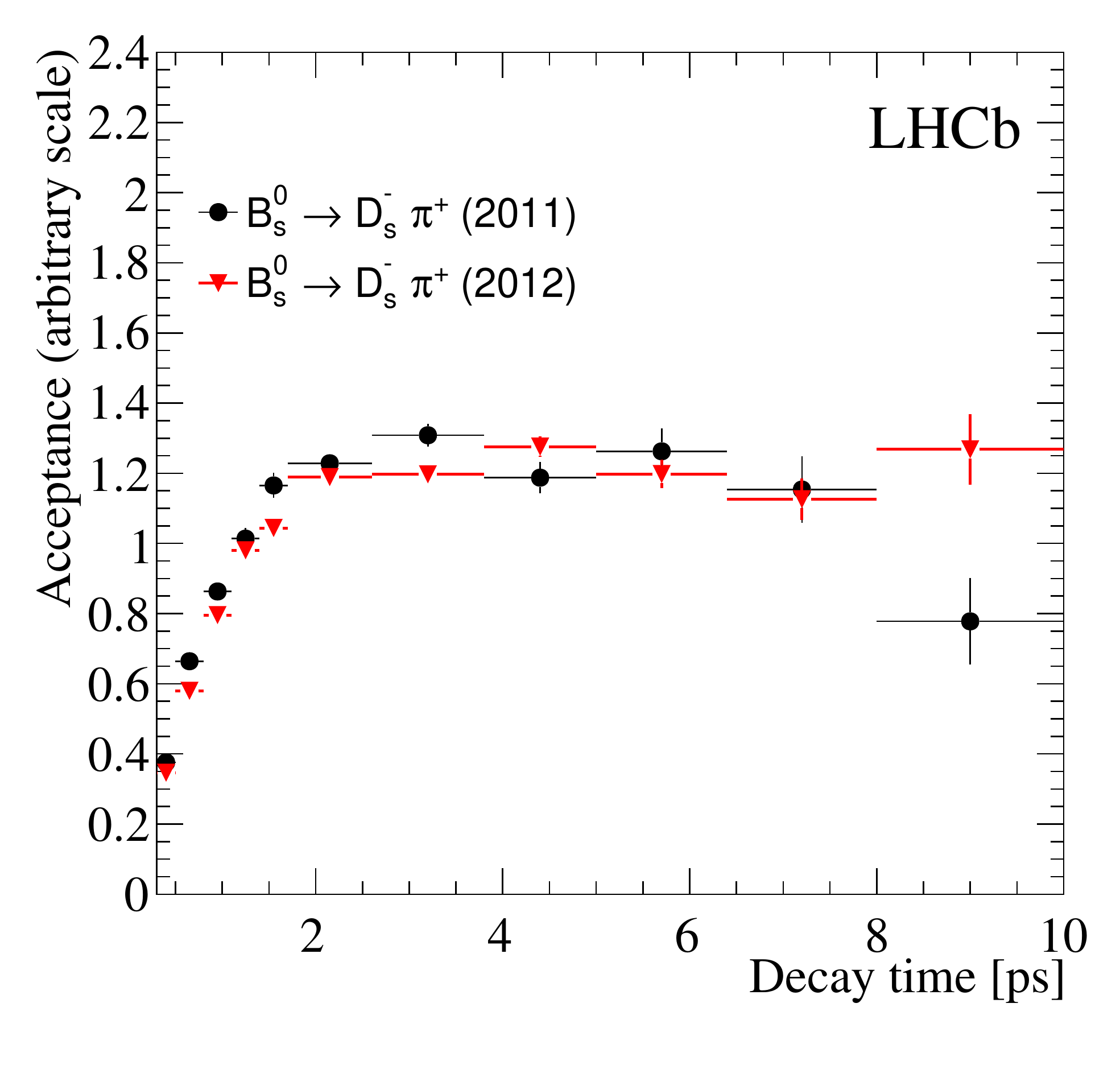}
\includegraphics[width=0.45\textwidth]{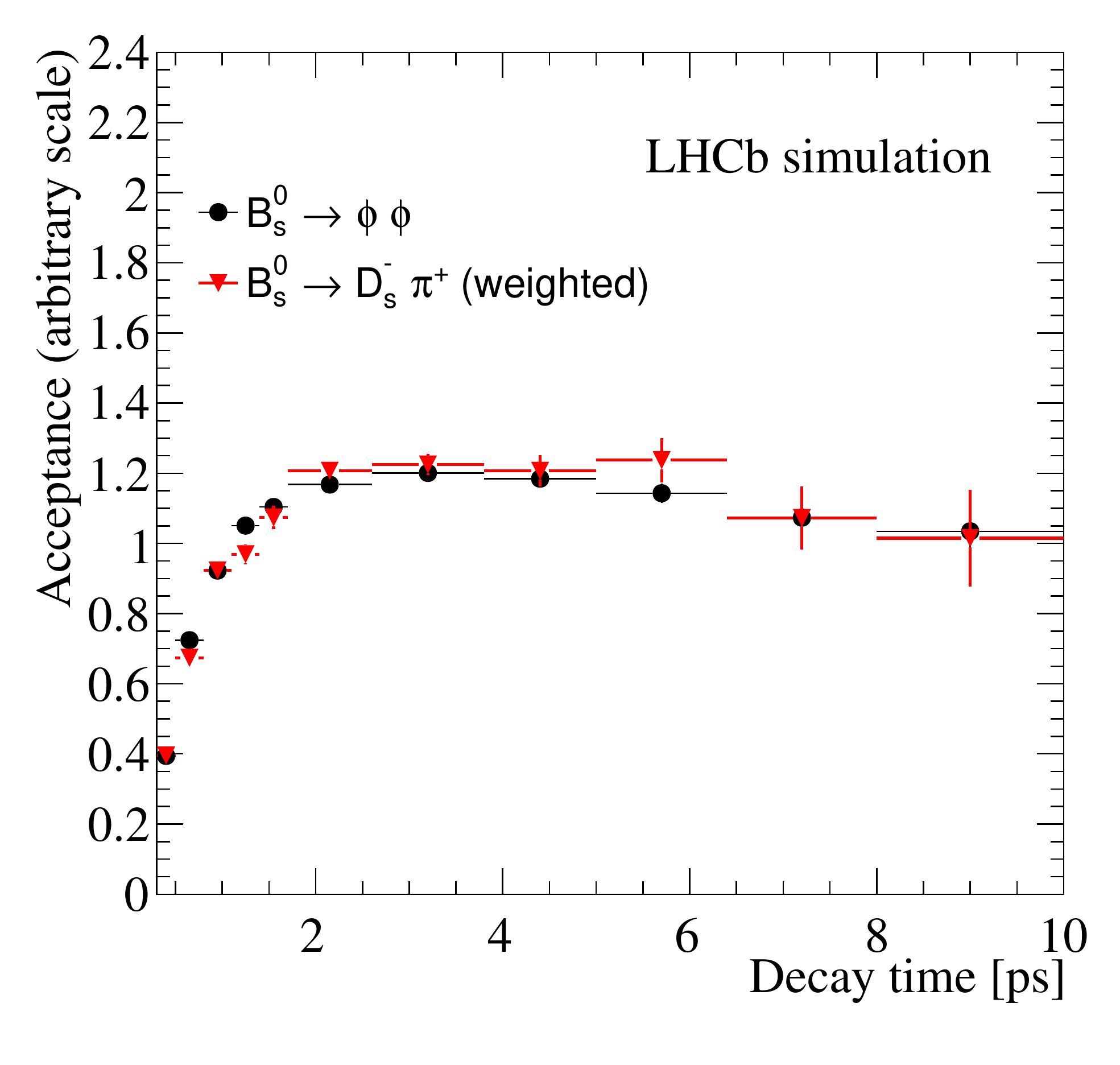}
\caption{
\small Decay time acceptance (left) calculated using $\Bs\to\Dsm\pip$ data events, and (right)
comparing \BsPP and $\Bs\to\Dsm\pip$ simulation, where $\Bs\to\Dsm\pip$ events
are re-weighted to match the distribution of the minimum \pt of the final state particles
in \BsPP decays.}
\label{fig:TA}
\end{figure}

For the case of the decay time dependent fit, the efficiency as a function of the decay time
is modelled as a histogram, with systematic uncertainties arising from the 
differences in \BsPP and $\Bs\to\Dsm\pip$ simulated events.
Figure~\ref{fig:TA} shows the comparison of the efficiency as a function of decay time calculated
using $\Bs\to\Dsm\pip$ data in 2011 and 2012. Also shown is the comparison between \BsPP and $\Bs\to\Dsm\pip$
simulated events.

In the fit to determine the triple-product asymmetries, the decay time acceptance
is treated only as a systematic uncertainty, which is based on
the acceptance found from $\Bs\to\Dsm\pip$ data events.

\section{Flavour tagging}
\label{sec:flavour}

To maximise the sensitivity on \phisPP, the 
determination of the initial flavour of the \Bs meson is necessary.
This results from the terms in the differential decay rate with the largest sensitivity to \phisPP
requiring the identification (tagging) of the flavour at production.
At \lhcb, tagging is achieved through the use of different algorithms
described in Refs.\cite{LHCb-PAPER-2013-002,LHCb-CONF-2012-033}. 
This analysis uses both the opposite side (OS) and same side kaon (SSK) flavour taggers. 

The OS flavour tagging algorithm~\cite{LHCb-CONF-2012-026} makes use of the $\bquarkbar(\bquark)$-quark 
produced in association with the signal $\bquark(\bquarkbar)$-quark. In this analysis, the predicted 
probability of an incorrect flavour assignment, $\omega$, is determined for each event by a neural network that is calibrated using $\Bu \to \jpsi \Kp$, 
$\Bu\to\Dzb\pip$, $\Bd\to\jpsi\Kstarz$, $\Bd\to\Dstarm\mup\neum$, and $\Bs \to \Dsm\pip$ data as control modes.
Details of the calibration procedure can be
found in Ref.~\cite{LHCb-PAPER-2013-002}. 

When a signal \Bs meson is formed, there is an associated \squarkbar quark formed in the first branches of the fragmentation that about 50\,\% of the time
forms a charged kaon, which is likely to originate close to the \Bs meson production point. The kaon charge therefore allows for the identification of the flavour of the signal \Bs meson.
This principle is exploited by the SSK flavour tagging algorithm~\cite{LHCb-CONF-2012-033}.
The SSK tagger is calibrated with the $\Bs \to \Dsp\pim$ decay mode.
A neural network is used to select fragmentation particles, 
improving the flavour tagging power quoted in the previous decay time dependent measurement~\cite{LHCb-PAPER-2013-007,Krocker:1631104}.

Flavour tagging power is defined as $\epsilon_{\rm tag} \mathcal{D}^2$, where $\epsilon_{\rm tag}$ is the
flavour tagging efficiency and $\mathcal{D}\equiv (1-2\omega)$ is the dilution.
Table~\ref{tab:stats} shows the tagging power for the events tagged by only one of the algorithms and those tagged by both,
estimated from 2011 and 2012 \BsPP data events separately.
Uncertainties due to the calibration of the flavour tagging algorithms are applied as Gaussian
constraints in the decay time dependent fit. The dependence of the flavour tagging 
initial flavour of the \Bs meson is accounted for during fitting.
\begin{table}[t]
\begin{center}
\caption{\label{tab:stats} \small Tagging efficiency ($\epsilon_{\rm tag}$), effective dilution ($\mathcal{D}$),
and tagging power ($\epsilon \mathcal{D}^2$), as estimated from the data
for events tagged containing information
from OS algorithms only, SSK algorithms only, and information from both algorithms.
Quoted uncertainties include both statistical and systematic contributions.}
\begin{tabular}{lrrr}
Dataset & $\epsilon_{\rm tag} \,(\%)$ & $\mathcal{D} \,(\%)$ & $\epsilon \mathcal{D}^2 \,(\%)$ \\ \hline
2011 OS& $12.3\pm1.0$ & $31.6\pm0.2$ & $1.23\pm0.10$ \\
2012 OS& $14.5\pm0.7$ & $32.7\pm0.3$ & $1.55\pm0.08$ \\\hline
2011 SSK& $40.2\pm1.4$ & $15.2\pm2.0$ & $0.93\pm0.25$ \\
2012 SSK& $33.1\pm0.9$ & $16.0\pm1.6$ & $0.85\pm0.17$ \\\hline
2011 Both& $26.0\pm1.3$ & $34.9\pm1.1$ & $3.17\pm0.26$ \\
2012 Both& $27.5\pm0.9$ & $33.2\pm1.2$ & $3.04\pm0.24$ \\
\end{tabular}
\end{center}
\end{table}

\section{Decay time dependent measurement}

\subsection{Likelihood}
\label{sec:LL_TD}

The parameters of interest are the \CP violation parameters ($\phisPP$ and $|\lambda|$), the polarisation amplitudes ($|A_0|^2$, $|A_\perp|^2$, $|A_S|^2$, and $|A_{SS}|^2$), 
and the strong phases ($\delta_1$, $\delta_2$, $\delta_S$, and $\delta_{SS}$), as
defined in Sec.~\ref{sec:modelTD}. The $P$-wave amplitudes are defined such that
$|A_0|^2+|A_\perp|^2+|A_\parallel|^2=1$, hence only two are free parameters.

Parameter estimation is achieved from a minimisation of the negative log likelihood.
The likelihood, $\mathcal{L}$, is weighted using the \sPlot method~\cite{Pivk:2004ty,Xie:2009rka},
with the signal weight of an event $e$ calculated from the equation
\begin{align}
W_e(m_{\Kp\Km\Kp\Km}) = \frac{\sum_{j} V_{sj}F_j(m_{\Kp\Km\Kp\Km})}{\sum_j N_jF_j(m_{\Kp\Km\Kp\Km})},
\end{align}
where $j$ sums over the number of fit components to the four-kaon invariant mass, with PDFs $F$,
associated yields $N$, and $V_{sj}$ is the covariance between the signal yield and the yield
associated with the $j^{\rm th}$ fit component.
The log-likelihood then takes the form
\begin{align}
-\ln \mathcal{L} = -\alpha \sum_{{\rm events}\, e} W_e \ln (S^e_{\rm TD}),
\end{align}
where $\alpha=\sum_eW_e/\sum_eW_e^2$ is used to account for the weights in the determination of the statistical uncertainties,
and $S_{\rm TD}$ is the signal model of Eq.~\ref{eq:pdf}, accounting also for the effects of decay time and
angular acceptance, in addition to the probability of an incorrect flavour tag.
Explicitly, this can be written as
\begin{align}
S^e_{\rm TD} = \frac{\sum_is^e_i(t_e)f_i(\Omega_e)\epsilon(t_e)}{\sum_k\zeta_k\int s_k(t)f_k(\Omega)\epsilon(t) \deriv t \, \deriv\Omega},
\end{align}
where $\zeta_k$ are the normalisation integrals used to describe the angular acceptance described in Sec.~\ref{sec:AA} and
\begin{align}
s^e_i(t)&=N_ie^{-\Gs t_e} \left[ c_iq_e(1-2\omega_e) \cos(\dms t_e) + d_iq_e(1-2\omega_e)\sin(\dms t_e) + a_i \cosh\left(\frac{1}{2}\DGs t_e\right) \right.
\nonumber\\ &\qquad\left. + b_i \sinh\left(\frac{1}{2}\DGs t_e\right)\right] \otimes R(\sigma^{\rm cal}_e,t_e),
\label{eq:explicit}
\end{align}
where $\omega_e$ is the calibrated probability of an incorrect flavour assignment, and $R$ denotes the Gaussian resolution function.
In Eq.~\ref{eq:explicit}, $q_e=1(-1)$ for a \Bs (\Bsb) meson at $t=0$ in event $e$ or $q_e=0$ if no flavour tagging information exists.
The 2011 and 2012 data samples are assigned independent signal weights, decay time and angular acceptances,
in addition to separate Gaussian constraints to the decay time resolution parameters as defined in Sec.~\ref{sec:DTR}.
The value of the \Bs-\Bsb oscillation frequency is constrained to the \lhcb measured value
of $\dms=17.768\pm0.023\stat\pm0.006\syst\invps$~\cite{LHCb-PAPER-2013-006}. The values of the decay
width and decay width difference are constrained to the \lhcb measured values
of $\Gs=0.661\pm0.004\stat\pm0.006\syst$\invps and $\DGs=0.106\pm0.011\stat\pm0.007\syst$\invps, respectively~\cite{LHCb-PAPER-2013-002}.
The Gaussian constraints applied to the \Gs and \DGs parameters use the combination of the measured
values from $\Bs\to\jpsi\Kp\Km$ and $\Bs\to\jpsi\pip\pim$ decays. Constraints are therefore applied
taking into account a correlation of $0.1$ for the statistical uncertainties~\cite{LHCb-PAPER-2013-002}.
The systematic uncertainties are taken to be uncorrelated between the $\Bs\to\jpsi\Kp\Km$ and $\Bs\to\jpsi\pip\pim$ decay modes.

The events selected in this analysis are within the two-kaon invariant mass range ${994.5 < m_{\Kp\Km} < 1044.5\mevcc}$,
and are divided into three regions. These correspond to both $\phi$ candidates
with invariant masses smaller than the known $\phi$ mass, one $\phi$ candidate with an invariant mass
smaller than the known $\phi$ mass and one larger, and a third region in which both $\phi$ candidates have
invariant masses larger than the known $\phi$ mass.
Binning the data in this way allows the analysis to become insensitive to correction
factors that must be applied to each of the $S$-wave and double $S$-wave interference terms in the differential cross section.
These factors modulate the contributions of the interference terms in the angular PDF due to the 
different line-shapes of kaon pairs originating from spin-1 and spin-0 configurations. 
Their parameterisations are denoted by $g(m_{\Kp\Km})$ and $h(m_{\Kp\Km})$, respectively. The spin-1
configuration is described by a Breit-Wigner function and the spin-0 configuration is assumed to be approximately uniform.  
The correction factors, denoted by $C_{SP}$, are defined from the relation~\cite{LHCb-PAPER-2013-002}
\begin{align}
C_{SP}e^{i\theta_{SP}} = \int^{m_{\rm h}}_{m_{\rm l}} g^*(m_{\Kp\Km}) h(m_{\Kp\Km}) \deriv m_{\Kp\Km}, 
\end{align}
where $m_{\rm h}$ and $m_{\rm l}$ are the upper and lower edges of a given $m_{\Kp\Km}$ bin, respectively.
Alternative assumptions on the $P$-wave and $S$-wave lineshapes are found to
have a negligible effect on the parameter estimation.

A simultaneous fit is then performed in the three $m_{\Kp\Km}$ invariant mass regions,
with all parameters shared except for the fractions and strong phases associated with the
$S$-wave and double $S$-wave, which are allowed to vary independently in each region.
The correction factors are calculated as described in Ref.~\cite{LHCb-PAPER-2013-002}.
The correction factor used for each region is calculated to be 0.69.

\subsection{Results}
\label{sec:resTD}

The results of the fit to the parameters of interest
are given in Table~\ref{tab:nominal}. 
\begin{table}[t]
\caption{\small \label{tab:nominal} Results of the decay time dependent fit.}
\begin{center}
\begin{tabular}{@{}lr}
Parameter & Best fit value \\ \hline
$\phisPP$ (\rad)	&      $-0.17\pm0.15$ 		\\
$|\lambda|$ 		&      $1.04\pm0.07$ 		\\\hline
$|A_\perp|^{2}$ 	&      $0.305 \pm    0.013$ 	\\
${|A_0|}^2$ 		&      $0.364 \pm    0.012$ 	\\
$\delta_1$ (\rad) 	&      $0.13 \pm     0.23$ 	\\
$\delta_2$ (\rad) 	&      $2.67 \pm    0.23$ 	\\\hline
\Gs (\invps) 		&      $0.662 \pm  0.006$ 	\\
\DGs (\invps) 		&      $0.102 \pm   0.012$ 	\\
$\dms$ (\invps) 	&      $17.774 \pm   0.024$ 	\\
\end{tabular}
\end{center}
\end{table}
The $S$-wave and double $S$-wave parameter estimations for the three regions
defined in Sec.~\ref{sec:LL_TD} are given in Table~\ref{tab:nominalS}.
The fraction of $S$-wave is found to be consistent with zero
in all three mass regions.
\begin{table}[t]
\begin{center}
\caption{\small \label{tab:nominalS} $S$-wave and double $S$-wave results of the decay time dependent fit 
for the three regions identified in Sec.~\ref{sec:LL_TD}, where $M_{--}$ indicates the
region with both two-kaon invariant masses smaller than the known $\phi$ mass, $M_{-+}$ the region with
one smaller and one larger, and $M_{++}$ indicates the region with both two-kaon invariant masses
larger than the known $\phi$ mass.
}
\begin{tabular}{l|rrrr}
Region & ${|A_S|}^2$ & $\delta_S$(\rad) & ${|A_{SS}|}^2$ & $\delta_{SS}$(\rad) \\ \hline
$M_{--}$ &      $0.006 \pm 0.012$ &    $-0.40  \pm 0.53$ &       $0.009 \pm 0.016$ &  $-2.99  \pm 1.27$ \\
$M_{-+}$ &      $0.006 \pm 0.010$ &    $2.76  \pm 0.39$ &        $0.004 \pm 0.011$ &  $-2.17  \pm 0.72$ \\
$M_{++}$ &      $0.001 \pm 0.003$ &    $-2.58  \pm 2.08$  &      $0.020 \pm 0.022$ &  $0.53  \pm 0.55$
\end{tabular}
\end{center}
\end{table}

The correlation matrix is shown in Table~\ref{tab:corr}. 
The largest correlations are found to be between the amplitudes themselves and the \CP-conserving strong phases themselves.
The observed correlations have been verified with simulated datasets.
Cross-checks are performed on simulated datasets generated with the same number of events as observed in data,
and with the same physics parameters, to ensure that generation values are recovered with negligible biases.
\begin{table}[t]
\caption{\small \label{tab:corr} Correlation matrix associated with the result of the decay time dependent fit.
Correlations with a magnitude greater than 0.5 are shown in bold.}
\begin{center}
{ \scriptsize
\begin{tabular}{l|rrrrrrrrrr}
 & $|A_{\perp}|^{2}$ & $|A_0|^2$ & $|A_{SS}|^2$ & $|A_S|^2$ & $\delta_{SS}$  & $\delta_S$ & $\delta_1$ & $\delta_2$ & $\phisPP$ & $|\lambda|$ \\ \hline
$|A_\perp|^{2}$   & 1.00 & --0.48 & 0.01 & 0.07 & 0.00 & 0.01 & --0.04 & 0.01 & --0.13 & --0.01\\
$|A_0|^2$ &  &   1.00 & --0.02 & --0.14 & --0.03 & 0.01 & 0.05 & 0.02 &  0.07 & 0.03  \\
$|A_{SS}|^2$ &    &  & 1.00 & 0.18 & \bf{0.59} & 0.01 & 0.04 & 0.07 & --0.03 & --0.18 \\
$|A_S|^2$ &  &    &  & 1.00 & 0.21 & 0.01 & 0.01 & 0.06 &  --0.03 & --0.25\\           
$\delta_{SS}$ &   &  &  &  & 1.00 & --0.02 & 0.03 & 0.06 &  --0.06 & --0.21 \\
$\delta_S$ &  &   &  &  &  & 1.00 & 0.40 & 0.42 & --0.07 & --0.16 \\
$\delta_1$ &  &   &  &  &  &  & 1.00 & \bf{0.95} & --0.20 & --0.27 \\
$\delta_2$ &  &   &  &  &  &  &  & 1.00 & --0.20 & --0.28 \\
$\phisPP$ &  &   &  &  &  &  &  &  &   1.00 & 0.12 \\
$|\lambda|$ &  &   &  &  &  &  &  &  &  & 1.00 \\
\end{tabular}
}
\end{center}
\end{table}

\begin{figure}[t]
\begin{center}
\includegraphics[height=5cm]{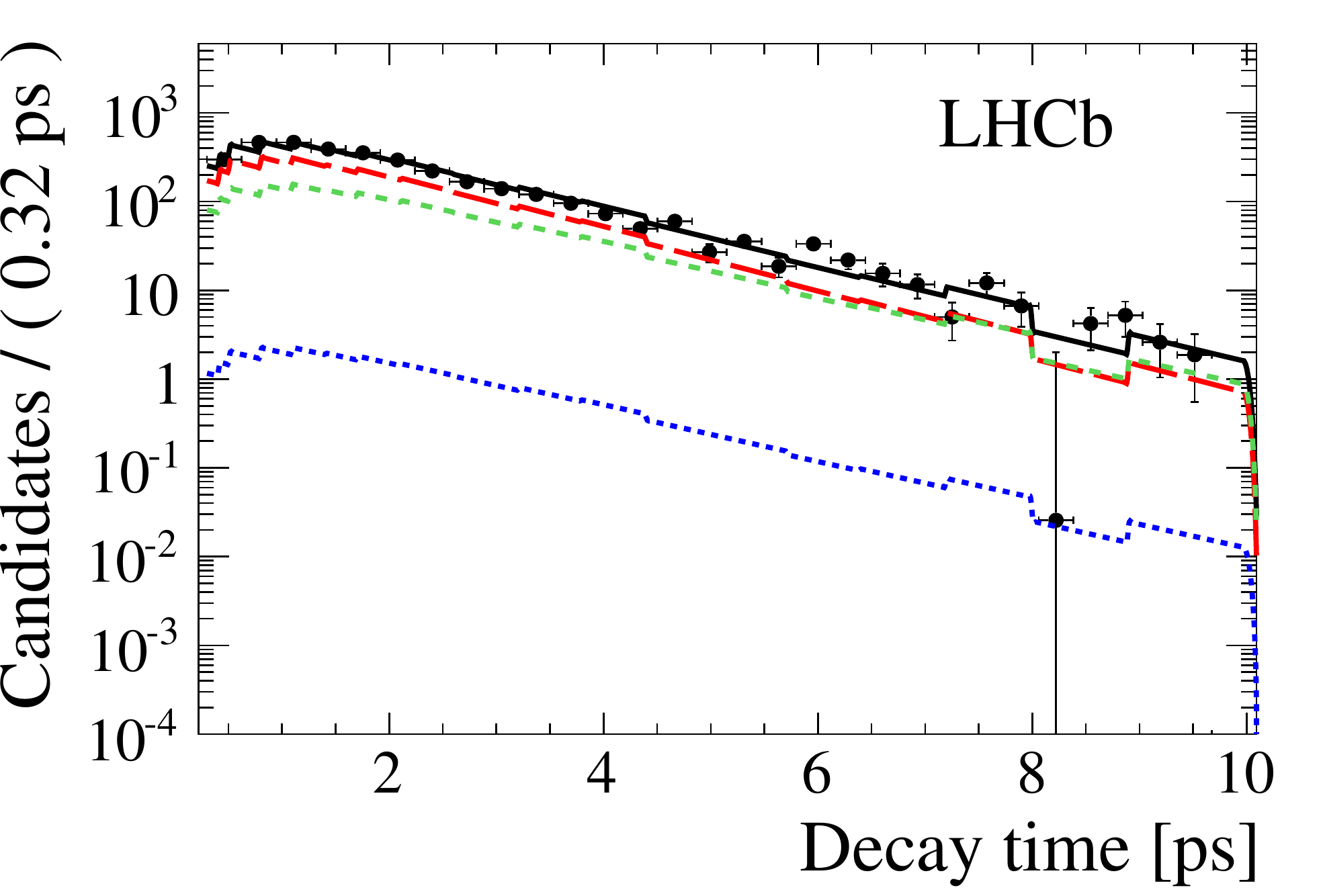}
\includegraphics[height=5cm]{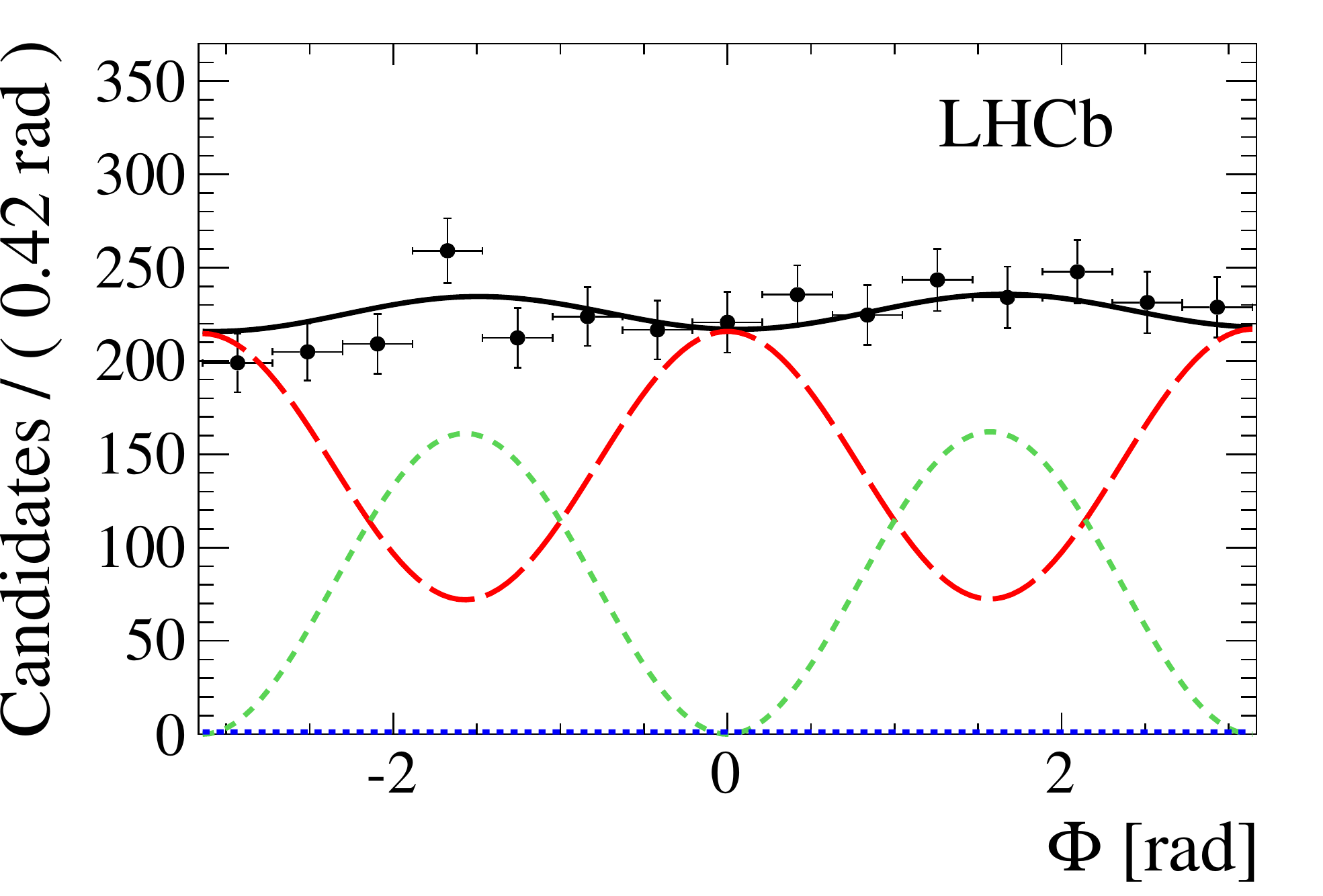}
\includegraphics[height=5cm]{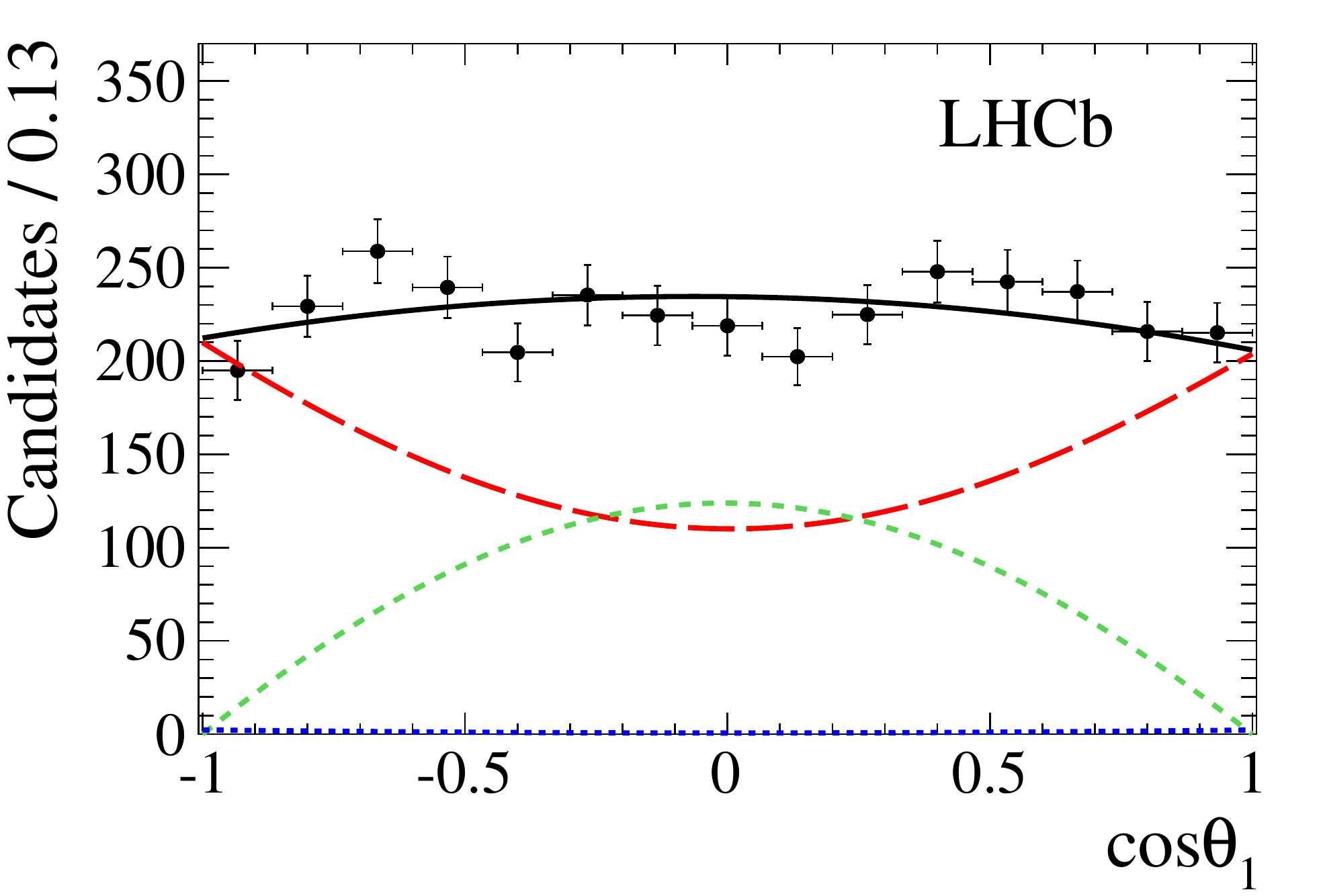}
\includegraphics[height=5cm]{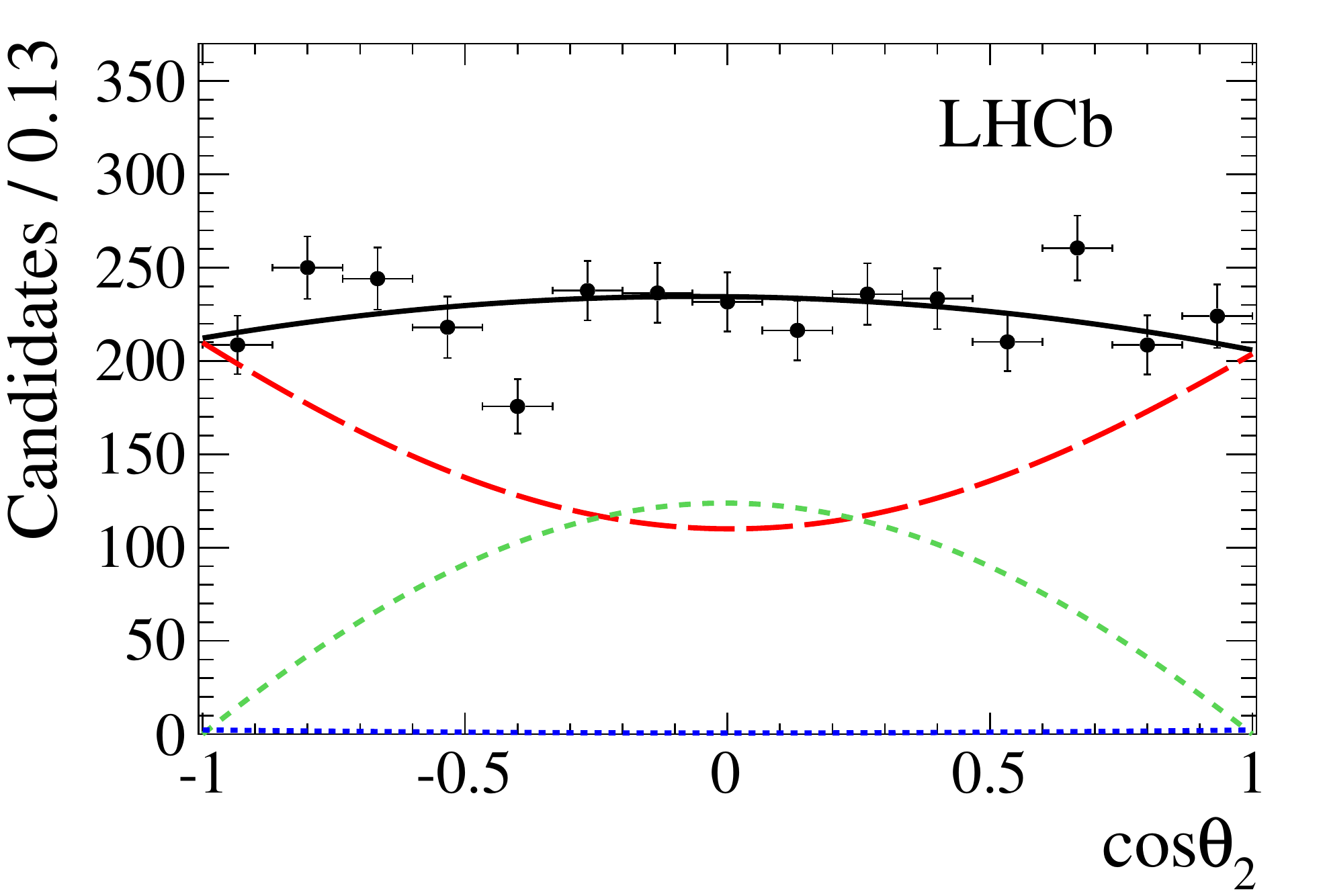}
\caption{One-dimensional projections of the $\Bs \to \phi\phi$ fit for (top-left) decay time with binned acceptance,
  (top-right) helicity angle $\Phi$ and (bottom-left and bottom-right) cosine
    of the helicity angles $\theta_1$ and $\theta_2$.
       The background-subtracted data are marked as black points, while the black solid lines represent the projections of the best fit.
          The \CP-even $P$-wave, the \CP-odd $P$-wave and $S$-wave combined with double $S$-wave
	     components are shown by the red long dashed, green short dashed and blue dotted lines, respectively.}
\label{fig:proj}
\end{center}
\end{figure}

Figure~\ref{fig:proj} shows the distributions of the \Bs decay time and the three
helicity angles.
Superimposed are the projections of the fit result. The projections are event weighted to yield the signal distribution
and include acceptance effects. 

\begin{figure}[t]
\begin{center}
\includegraphics[height=6cm]{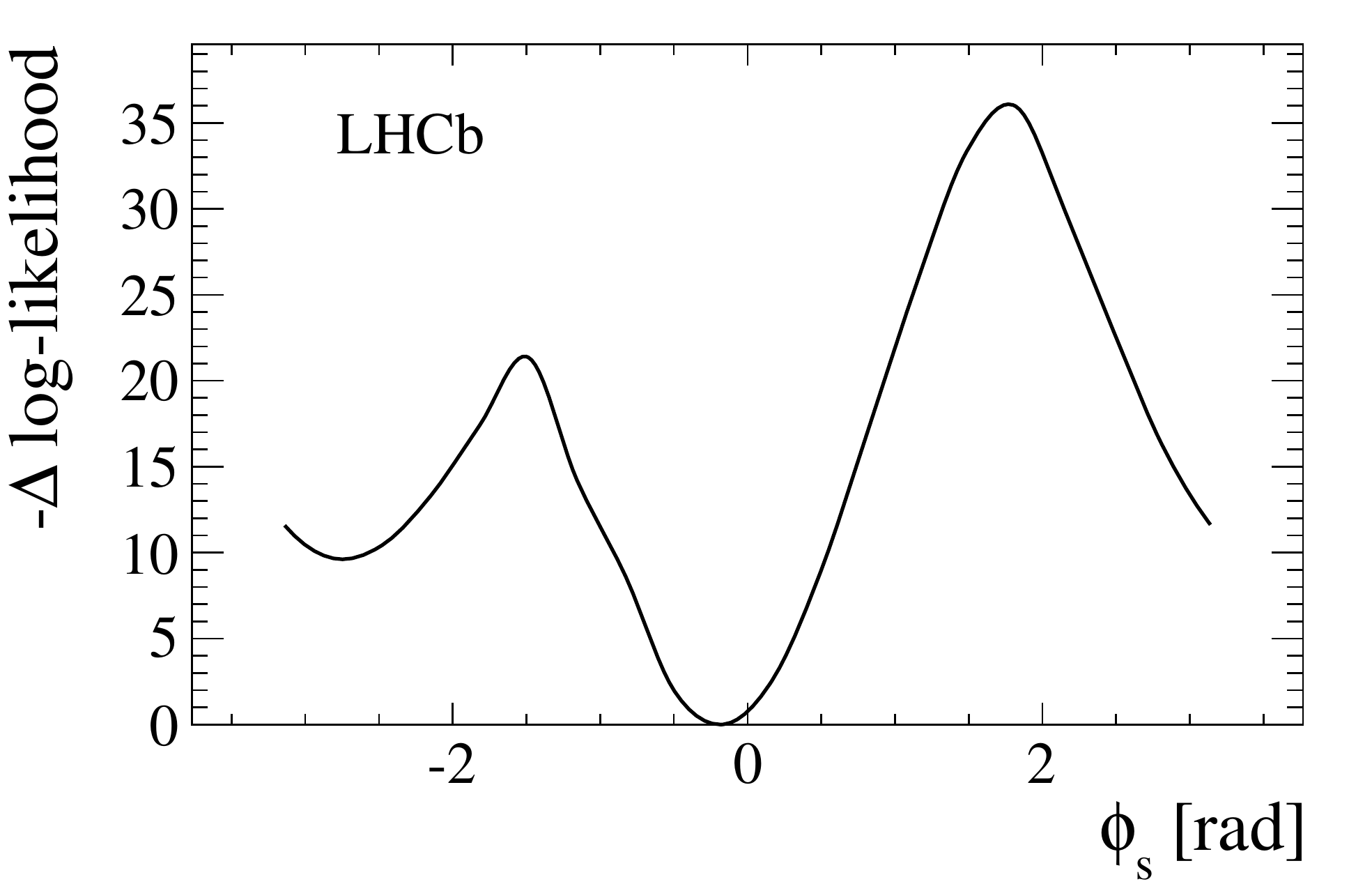}
\caption{Profile log-likelihood for the \phisPP parameter.}
\label{fig:phisLL}
\end{center}
\end{figure}

The scan of the natural logarithm of the likelihood for the \phisPP parameter is shown in Fig.~\ref{fig:phisLL}.
At each point in the scan, all other parameters are re-minimised.
A parabolic minimum is observed, and a point estimate provided.
The shape of the profile log-likelihood is replicated in simplified simulations as a cross-check.

\subsection{Systematic uncertainties}
\label{sec:syst}

The most significant systematic effects arise from the angular and decay time acceptances.
Minor contributions are also found from the mass model used to construct the event
weights, the uncertainty on the peaking background contributions, and 
the fit bias.

An uncertainty due to the angular acceptance arises from the limited number of simulated
events used to determine the acceptance correction. This is accounted for by varying
the normalisation weights within their statistical uncertainties accounting for correlations.
The varied weights are then used to fit simulated datasets. This process is repeated
and the width of the Gaussian distribution 
is used as the uncertainty. A further uncertainty arises from the
assumption that the angular acceptance does not depend on the algorithm
used for the initial flavour assignment. Such a dependence can be expected
due to the kinematic correlations of the tagging particles with the signal particles. This
introduces a tagging efficiency based on the kinematics of the signal particles.
The difference
between the nominal data result and the result with angular acceptances calculated independently for the
different flavour tagging algorithms leads to a non-negligible uncertainty on the polarisation
amplitudes. Further checks are performed to verify that the angular acceptance does not depend
on the way in which the event was triggered.

The systematic uncertainty on the decay time acceptance is evaluated
from the difference in the decay time acceptance evaluated from \BsPP and $\Bs\to\Dsm\pip$
simulated events. The simulated datasets are generated with the decay time acceptance of
\BsPP simulation and then fitted with the $\Bs\to\Dsm\pip$ decay time acceptance.
This process is repeated and the resulting bias on the fitted parameters
is used as an estimate of the systematic uncertainty.

The uncertainty on the mass model is found by refitting the data
with signal weights derived from a single Gaussian \BsPP model, rather than the
nominal double Gaussian. The uncertainty due to peaking background contributions
is found through the recalculation of the signal weights with peaking background
contributions varied according to the statistical uncertainties on the yields
of the $\Lb\to\phi \proton\Km$ and $\Bd\to\phi\Kstarz$ contributions. Fit
bias arises in likelihood fits when the number of events used to determine the free parameters
is not sufficient to achieve the Gaussian limit. This uncertainty is evaluated by generating and fitting 
simulated datasets and taking the resulting bias as the uncertainty.

Uncertainties due to flavour tagging are included in the statistical uncertainty through
Gaussian constraints on the calibration parameters, and amount to 10\,\% of the statistical
uncertainty on the \CP-violating phase.

A summary of the systematic uncertainties is given in Table~\ref{tab:syst_sum}. 
\begin{table}[t]
\begin{center}
\caption{\small Summary of systematic uncertainties for physics parameters in the decay time dependent measurement,
where AA denotes angular acceptance.}
\begin{tabular}{lrrrrrr}%p{3.5cm}RRRSSS
Parameter      & $|A_0|^2$ & $|A_\perp|^2$    & $\delta_1$ (rad)     & $\delta_2$ (rad)    & \phisPP (rad)     & $|\lambda|$ \\ \hline
Mass model  & --      & --      & 0.03        & 0.04     & --        & 0.02    \\
AA (statistical) & 0.003        & 0.004      & 0.02        & 0.02        & 0.02        & 0.02      \\
AA (tagging)& 0.006      & 0.002       & --        & 0.01      & --         & 0.01     \\
Fit bias & --      &--       & 0.02        & --        &--         & --     \\
Time acceptance    & 0.005       & 0.003       & 0.02        & 0.05        & 0.02         & --       \\
Peaking background & --      & --       & 0.01        & 0.01        & --        & 0.01   \\
Total & 0.009      & 0.005       & 0.05        & 0.07        & 0.03  & 0.03     \\
\end{tabular}
\label{tab:syst_sum}
\end{center}
\end{table}

\section{Triple-product asymmetries}
\subsection{Likelihood}
\label{sec:LL_TPA}

In order to determine the triple-product asymmetries, a separate likelihood fit is performed. 
This is based around the simultaneous fitting of separate datasets to the four-kaon invariant mass,
which are split according to the sign of $U$ and $V$ observables.
Simultaneous mass fits are performed for the $U$ and $V$ observables separately.
The set of free parameters in fits to determine the $U$ and $V$ observables consist of the asymmetries of the \BsPP signal and
combinatoric background ($A_{U(V)}$ and $A_{U(V)}^{\rm B}$), along with their associated total yields ($N_{\rm S}$ and $N_{\rm B}$).
The mass model is the same as that described in Sec.~\ref{sec:selection}.
The total PDF, $S_{\rm TP}$ is then of the form
\begin{equation}
S_{\rm TP} = \sum_{i \in \{+,-\}} \left( f_i^S G^S(m_{\Kp\Km\Kp\Km}) + \sum_{j}f^j_i P^j(m_{\Kp\Km\Kp\Km}) \right),
\label{eq:newtotpdf}
\end{equation}
where $j$ indicates the sum over the background components with corresponding PDFs, $P^j$,
and $G^S$ is the double Gaussian signal PDF as described in Sec.~\ref{sec:selection}.
The parameters $f^k_i$ found in Eq.~\ref{eq:newtotpdf} are related to the asymmetry, $A_{U(V)}^k$, through
\begin{eqnarray}
f_+^{k} = \frac{1}{2}(A^k_{U(V)} + 1), \\
f_-^{k} = \frac{1}{2}(1-A^k_{U(V)}),
\label{eq:asymmetries1}
\end{eqnarray}
where $k$ denotes a four-kaon mass fit component, as described in Sec.~\ref{sec:selection}. 
Peaking backgrounds are assumed to be symmetric in $U$ and $V$.

\subsection{Results}
\label{sec:resTP}

\begin{figure}[t]
\includegraphics[width=0.5\textwidth]{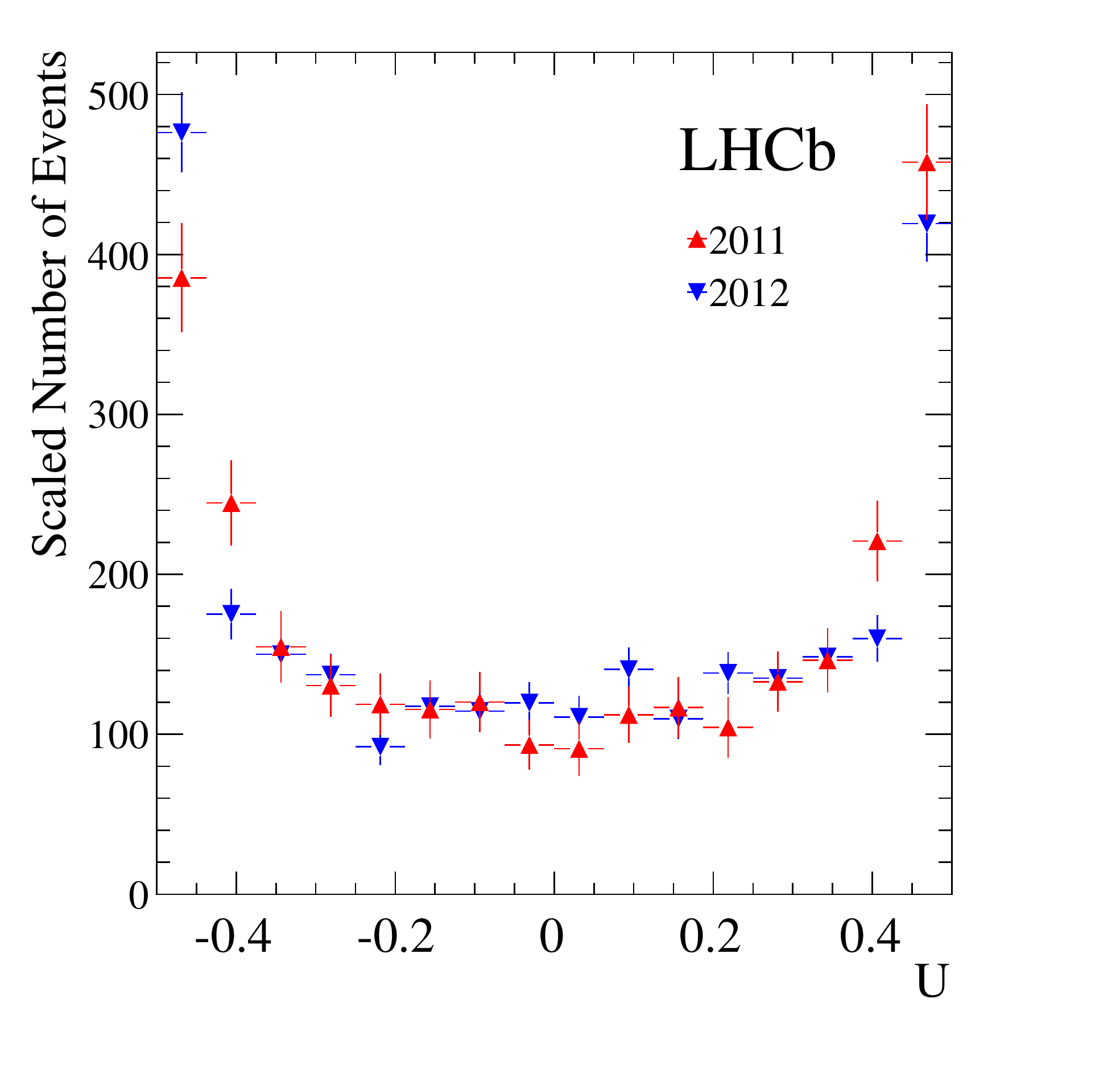}
\includegraphics[width=0.5\textwidth]{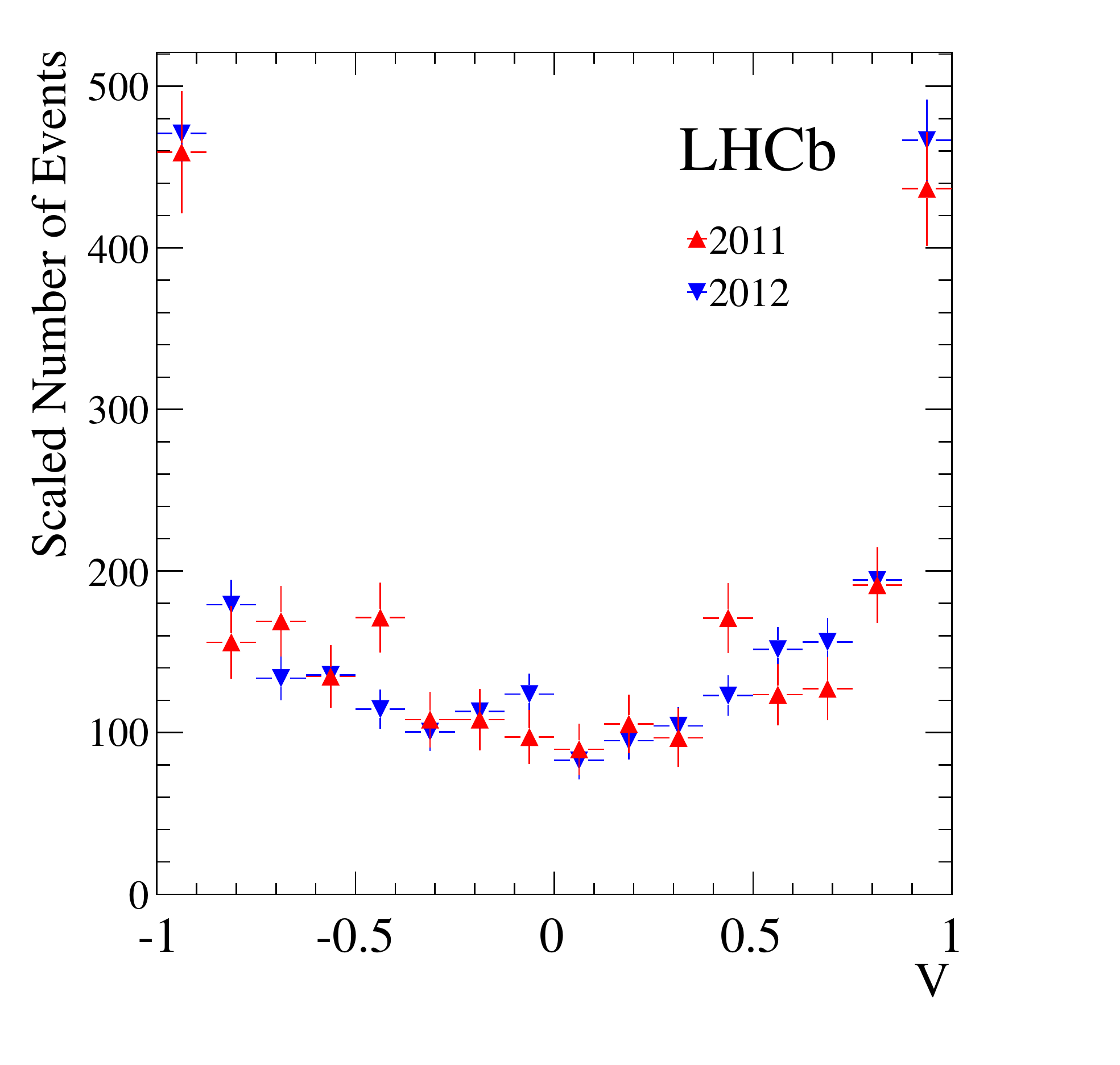}
\caption{Background-subtracted distributions of the (left) $U$ and (right) $V$ observables for the 2011 and 2012 datasets 
and restricted to the mass range $5246.8<m_{\Kp\Km\Kp\Km}<5486.8\mevcc$. The 2011 distributions are
scaled to have the same area as the 2012 distributions.
}
\label{UVobs}
\end{figure}

The background-subtracted distributions of the $U$ and $V$ observables are shown in
Fig.~\ref{UVobs} for the mass range  $5246.8<m_{\Kp\Km\Kp\Km}<5486.8\mevcc$. 
Distributions are found to agree between 2011 and 2012 datasets and show qualitatively symmetric
distributions.
The triple-product asymmetries found from the simultaneous fit described in Sec.~\ref{sec:LL_TPA} are measured to be
\begin{center}
\begin{tabular}{l@{$~=~$}r@{$\,\,\pm\,$}l}
$A_U$ & $-$0.003 & 0.017$,$  \\
$A_V$ & $-$0.017 & 0.017$.$
\end{tabular}
\end{center}

Statistical uncertainties are therefore to have approximately halved
with respect to the previous \lhcb measurements~\cite{LHCb-PAPER-2012-004}, due to more
efficient selection requirements and a larger data sample, and
are verified through fits to simulated datasets.
No evidence for \CP
violation is found.

\subsection{Systematic uncertainties}
\label{sec:systTP}

As for the case of the decay time dependent fit, the largest contributions to 
the systematic uncertainty arise from the decay time and angular acceptances.
Minor uncertainties also result from the mass model and peaking background knowledge.

The effect of the decay time acceptance is determined through the generation of
simulated samples including the decay time acceptance obtained from $\Bs\to\Dsm\pip$ data,
and fitted with the method described in Sec.~\ref{sec:LL_TPA}.
The resulting bias is used to assign a systematic uncertainty.

The effect of the angular acceptance is evaluated by generating simulated datasets with and without
the inclusion of the angular acceptance. The resulting bias found on the fit results of the 
triple-product asymmetries is then used as a systematic uncertainty.

Uncertainties related to the mass model are evaluated by
taking the difference between the nominal fit results and those
using a single Gaussian function to model the \BsPP decay. The effect
of the peaking background is evaluated by taking the largest difference
between the nominal fit results and the fit results with the peaking background
yields varied according to their uncertainties, as given in Sec.~\ref{sec:selection}.

The total systematic uncertainty is
estimated by choosing the larger of the two individual systematic uncertainties on  $A_U$ and
$A_V$. The contributions are combined
in quadrature to determine the total systematic uncertainty. 

Systematic uncertainties due to the residual effect of the decay time, 
geometrical acceptance and the signal and background fit models are summarised in
Table~\ref{tab:systTP}. 
\begin{table}[t]
\caption[]{  {\label{tab:systTP}}  Systematic
  uncertainties on the triple-product asymmetries $A_U$ and $A_V$. The total uncertainty is the sum in quadrature of the larger 
  of the two components for each source.}
\begin{center}
\begin{tabular}{lrrr}
Source & $A_U$ & $A_V$ & Uncertainty \\ \hline
Angular acceptance                   & 0.001 & 0.003 & 0.003  \\
Time acceptance                      & 0.005 & 0.003 & 0.005  \\
Mass model                    & 0.002 & 0.002 & 0.002  \\
Peaking background   & --      & 0.001 & 0.001 \\ \hline
Total                             & 0.006 & 0.005 & 0.006
\end{tabular}
\end{center}
\end{table}

\section{Summary and conclusions}
\label{sec:summary}

Measurements of \CP violation in the \BsPP decay are presented, based on the full \lhcb 
Run~1 dataset of 3.0\invfb. The \CP-violating phase, \phisPP, and \CP violation parameter, $|\lambda|$,
are determined to be
\begin{center}
\begin{tabular}{l@{$~=~$}r@{$\,\,\pm\,$}r@{$\,\,\pm\,$}l}
\phisPP & $-$0.17 & 0.15\stat & 0.03\syst$\rad,$ \\
$|\lambda|$ & 1.04 & 0.07\stat & 0.03\syst$\kern-0.25em .$
\end{tabular}
\end{center}
Results are found to agree with the theoretical predictions~\cite{Bartsch:2008ps,beneke,PhysRevD.80.114026}.
When compared with the \CP-violating phase measured in ${\Bs\to\jpsi\Kp\Km}$ and ${\Bs\to\jpsi\pip\pim}$
decays~\cite{LHCb-PAPER-2013-002}, these results show that no large \CP violation is present either in \Bs-\Bsb mixing or in the $\bquarkbar\to\squarkbar\ssbar$
decay amplitude.

The polarisation amplitudes and strong phases are measured to be
\begin{center}
\begin{tabular}{l@{$~=~$}r@{$\,\,\pm\,$}r@{$\,\,\pm\,$}l}
$|A_0|^2$ &         0.364 & 0.012\stat & 0.009\syst$\kern-0.25em ,$         \\
$|A_\perp|^2$ &     0.305 & 0.013\stat & 0.005\syst$\kern-0.25em ,$         \\
$\delta_1$ &        0.13 & 0.23\stat & 0.05\syst$\rad ,$        \\
$\delta_2$ &        2.67 & 0.23\stat & 0.07\syst$\rad .$         
\end{tabular}
\end{center}
Values of the polarisation amplitudes are found to agree well with the previous measurements~\cite{LHCb-PAPER-2012-004,LHCb-PAPER-2013-007,Aaltonen:2011rs}. 
Measurements in other $B \rightarrow VV$ penguin transitions at the
\textit{B} factories generally give higher values of $f_L\equiv|A_0|^2$~\cite{PhysRevLett.94.221804,PhysRevLett.98.051801,PhysRevD.78.092008, delAmoSanchez:2010mz,Abe:2004mq, Aubert:2006fs}. 
The value of $f_L$ found in the \BsPP channel is almost equal to that in
the $\Bs \rightarrow \Kstarz \Kstarzb$ decay~\cite{kstarlhcb}. As reported in Ref.~\cite{LHCb-PAPER-2012-004}, the 
results are in agreement with QCD factorisation predictions
\cite{beneke, PhysRevD.80.114026}, but disfavour the perturbative QCD estimate given in Ref.\cite{PhysRevD.76.074018}. 
The fractions of $S$-wave and double $S$-wave are found to be consistent with zero in all three regions
of $m_{\Kp\Km}$ mass.

The triple-product asymmetries are determined from a separate decay time integrated fit to be
\begin{center}
\begin{tabular}{l@{$~=~$}r@{$\,\,\pm\,$}r@{$\,\,\pm\,$}l}
$A_U$ & $-$0.003 & 0.017\stat & 0.006\syst$\kern-0.25em ,$  \\
$A_V$ & $-$0.017 & 0.017\stat & 0.006\syst$\kern-0.25em ,$
\end{tabular}
\end{center}
in agreement with previous measurements~\cite{LHCb-PAPER-2012-004,Aaltonen:2011rs}. 

The results of the polarisation amplitudes, strong phases, and triple-product asymmetries 
presented in this paper supersede the previous \lhcb measurements~\cite{LHCb-PAPER-2012-004,LHCb-PAPER-2013-007}. 
The measured values of the \CP-violating phase and triple-product asymmetries are consistent 
with the hypothesis of \CP conservation.

\section*{Acknowledgements}

\noindent We express our gratitude to our colleagues in the CERN
accelerator departments for the excellent performance of the LHC. We
thank the technical and administrative staff at the LHCb
institutes. We acknowledge support from CERN and from the national
agencies: CAPES, CNPq, FAPERJ and FINEP (Brazil); NSFC (China);
CNRS/IN2P3 (France); BMBF, DFG, HGF and MPG (Germany); SFI (Ireland); INFN (Italy);
FOM and NWO (The Netherlands); MNiSW and NCN (Poland); MEN/IFA (Romania);
MinES and FANO (Russia); MinECo (Spain); SNSF and SER (Switzerland);
NASU (Ukraine); STFC (United Kingdom); NSF (USA).
The Tier1 computing centres are supported by IN2P3 (France), KIT and BMBF
(Germany), INFN (Italy), NWO and SURF (The Netherlands), PIC (Spain), GridPP
(United Kingdom).
We are indebted to the communities behind the multiple open
source software packages on which we depend. We are also thankful for the
computing resources and the access to software R\&D tools provided by Yandex LLC (Russia).
Individual groups or members have received support from
EPLANET, Marie Sk\l{}odowska-Curie Actions and ERC (European Union),
Conseil g\'{e}n\'{e}ral de Haute-Savoie, Labex ENIGMASS and OCEVU,
R\'{e}gion Auvergne (France), RFBR (Russia), XuntaGal and GENCAT (Spain), Royal Society and Royal
Commission for the Exhibition of 1851 (United Kingdom).

\ifx\mcitethebibliography\mciteundefinedmacro
\PackageError{LHCb.bst}{mciteplus.sty has not been loaded}
{This bibstyle requires the use of the mciteplus package.}\fi
\providecommand{\href}[2]{#2}

\newpage
% Author List ----------------------------
%%%%%%%%%%%%%%%%%%%%%%%%%%%%%%%%%%%%%%%%%%
\centerline{\large\bf LHCb collaboration}
\begin{flushleft}
\small
R.~Aaij$^{41}$, 
B.~Adeva$^{37}$, 
M.~Adinolfi$^{46}$, 
A.~Affolder$^{52}$, 
Z.~Ajaltouni$^{5}$, 
S.~Akar$^{6}$, 
J.~Albrecht$^{9}$, 
F.~Alessio$^{38}$, 
M.~Alexander$^{51}$, 
S.~Ali$^{41}$
% ...

{\footnotesize \it
% ...
$ ^{5}$Clermont Universit\'{e}, Universit\'{e} Blaise Pascal, CNRS/IN2P3, LPC, Clermont-Ferrand, France\\
$ ^{6}$CPPM, Aix-Marseille Universit\'{e}, CNRS/IN2P3, Marseille, France\\
% ...
$ ^{9}$Fakult\"{a}t Physik, Technische Universit\"{a}t Dortmund, Dortmund, Germany\\
% ...
$ ^{37}$Universidad de Santiago de Compostela, Santiago de Compostela, Spain\\
$ ^{38}$European Organization for Nuclear Research (CERN), Geneva, Switzerland\\
% ...
$ ^{41}$Nikhef National Institute for Subatomic Physics, Amsterdam, The Netherlands\\
% ...
$ ^{46}$H.H. Wills Physics Laboratory, University of Bristol, Bristol, United Kingdom\\
% ...
$ ^{51}$School of Physics and Astronomy, University of Glasgow, Glasgow, United Kingdom\\
$ ^{52}$Oliver Lodge Laboratory, University of Liverpool, Liverpool, United Kingdom\\
% ...
}
\end{flushleft}
%%%%%%%%%%%%%%%%%%%%%%%%%%%%%%%%%%%%%%%%%%

\end{document}